\def\DraftSize{
 \documentstyle{article}
 \textwidth 172mm
 \textheight 240mm
 \topmargin -40pt
 \oddsidemargin -20pt
 \renewcommand{\baselinestretch}{0.95}
}

\def\FullSize{
 \documentstyle[12pt]{article}
 \textwidth17cm
 \textheight22cm
 \oddsidemargin0em
 \evensidemargin0em
 \topmargin-2ex
 \parindent=3em
}

\DraftSize

\makeatletter
\@addtoreset{equation}{section}

\makeatother

\def\beq{\begin{equation}}
\def\eeq{\end{equation}}
\def\bea{\begin{eqnarray}}
\def\nn{\nonumber \\ }
\def\eea{\end{eqnarray}}
\def\ds{\displaystyle}
\def\nsz{\normalsize}
\def\ssz{\scriptsize}

\def\ni{\noindent}

\def\req#1{(\ref{#1})}

\def\rot{{\rm rot}\ }
\def\div{{\rm div}\ }
\def\Tr{{\rm Tr}\ }
\def\rep{{\rm Re}\ }

\def\scapt#1{\caption{\protect\small #1}}
\def\barc{\begin{array}{c}}
\def\ear{\end{array}}

\renewcommand{\baselinestretch}{1.4}

\title{
Complete zeta-function approach to
the electromagnetic Casimir effect for spheres and circles. \\
\vspace{5mm}
\ssz accepted in Annals of Physics.
}
\author{
\nsz S. Leseduarte$^1$ \thanks{E-mail: lese@ecm.ub.es}
and August Romeo$^2$ \thanks{E-mail: august@ceab.es},
\\
\ssz 1) \it Dept ECM and IFAE, Faculty of Physics, University of Barcelona,
\ssz\it Diagonal 647, 08028 Barcelona \\
\ssz 2) \it Blanes Centre for Advanced Studies (CEAB), CSIC,
\ssz\it Cam{\'\i} de Santa B{\`a}rbara, 17300 Blanes
}

\date{}

\begin{document}

\maketitle
\begin{abstract}
A technique for evaluating the electromagnetic Casimir energy
in situations involving spherical or circular boundaries is presented.
Zeta function regularization is unambiguously used from the start
and the properties of Bessel and related zeta functions are central.
Nontrivial results concerning these functions are given.
While part of their application agrees with previous
knowledge, new results like the zeta-regularized electromagnetic
Casimir energy for a circular wire are included.
\end{abstract}


\section{Introduction}

Casimir effect problems have been
showing a remarkably long-lived appeal
since the day of their birth\cite{Cas},
still stirring up intense activity all along the eighties
(\cite{To}-\cite{BVW}, to name just a few works),
and reaching present-day topics (e.g. \cite{BrSkSo}-\cite{Ne}).
During all this time they have been object of many different approaches:
stress-energy tensor \cite{DeCa},
Green function methods \cite{MiRaSc},
multiple scattering expansions \cite{BaDu},
heat-kernel series (\cite{SeGi}-\cite{Do}), etc.

In the present paper, we offer new calculations of
the Casimir energy for an e.m. field in the presence of
a sphere in $D=3$, and of a circle in $D=2$,
with perfect-conductor conditions at their points.
The contributions
from inside and outside the boundary are separately studied.
Further, in $D=3$ we take apart the pieces associated to
trasnverse electric (TE) and transverse magnetic (TM) modes in each case.
Although this is lengthier than
starting from the whole sum (since some pieces which cancel would
then be eliminated at the beginning, and now we keep them until
the end) we get as a bonus a useful decomposition into the four
contributions. Thus, we find as a byproduct the effect coming
from e.g. the interior part only or the Dirichlet modes by themselves.

Unlike previous works on this subject involving zeta-functions
at some stage or other (e.g. \cite{Mi,BeMi}),
we adopt here what might be called a `straight' or `frontal'
zeta-regularization approach right at the outset
from the eigenmode summations themselves,
as advocated e.g. in \cite{BVW}

We believe that our attitude is quite sound, as
generic zeta function regularization \cite{BMDC,Haw} is already a
widespread technique.
This
method, often applied when boundary conditions
affect a given quantum system ---like quantum billiards \cite{BeRo,It}---
or field theory, is also adequate
for dealing with finite temperature or curved spaces. Moreover, its
variants have suceeded in problems involving nontrivial topologies or
boundaries, like Kaluza-Klein style models \cite{Dow,CW},
or systems under the influence of external fields \cite{GV} or
in interaction with material bodies (e.g. superconductors \cite{KuUe})
---other applications are shown in \cite{EORBZ}.

Many of these advances came true by means of a
precise knowledge of the spectral zeta function associated to each
system, which is the key to the derivation of many results.
Somehow, this happens also to the present paper: our calculation is
indebted to the investigation started in \cite{ELR,LR} about Bessel
zeta functions\cite{Math,Ste},
and taken further in \cite{RU} for the case of a purely
scalar field inside a sphere, under Dirichlet boundary conditions.
That work is here extended so as to
include all the field modes appearing in the problems under
discussion.
It is also worth to note that a similar technique has been applied
in Quantum Cosmology to study the Hartle-Hawking wave
function of the universe \cite{Bar}.

After a brief survey on electromagnetism (sect. 2)
general considerations about the physical problem of the
Maxwell modes, and its
spectral zeta function are made in sect. 3.
First, we study in some detail the part associated to the
internal transverse electric (TE) modes with particular
emphasis on the zeta function for the zeros of the
$J_{\nu}$ Bessel function and on the construction
of the corresponding complete zeta function.
Then, the same procedure is applied to the internal transverse
magnetic (TM) part and its analogous zeta function.
Next, the method is repeated for external modes in sect. 4.
Comments about the result in these and related situations are made
in sect. 5.
Essential mathematical material concerning the derivation of partial
wave zeta functions for external modes has been packaged into app. A.
App. B contains an example of
complete zeta function alternative calculation by using the coefficients
of the heat-kernel series for the Laplacian.
A comparison with previous estimates for $D=2$ is made in app. C.

\section{Electromagnetism in $D=3,2$}

\subsection{Neutral and perfectly conducting sphere in $D=3$}

We shall briefly sketch the classical problem of
an electromagnetic (e.m.) field kept within a cavity resonator bounded
by a perfectly conducting spherical shell of radius $a$.
The adequate conditions on the surface are
$
\vec{n} \cdot \vec{B}|_{r=a}= \vec{n}\times \vec{E}|_{r=a}=0
$
(where $\vec{n}$ is the normal vector),
in addition to the requirement of regularity in the interior.
Then, the spherical e.m. waves which result from
solving Maxwell's equations have radial parts
$\propto r^{1-D/2} J_{\nu(D,l)} (\omega r)$, where
\beq
\ds\nu(D,l) = \ds l+{D \over 2}-1
\label{defnul}
\eeq
denotes the Bessel index for angular monentum $l$
in $D=3$ space dimensions.
In the present cirsumstances, the solutions are divided
into `transverse electric' (TE) and `transverse magnetic' (TM) ones
(see e.g. \cite{SlFr,Bo}).
Their possible associated frequencies ($\omega$) are then determined
by conditions on the radial parts which read, in each case,
\begin{eqnarray}
\ds \left. r^{1-D/2}J_{\nu(D,l)}( \omega r) \right\vert_{r=a}&=&0,
\mbox{ for region I  TE-modes}, \label{condTEI} \\
\ds \left. {d \over dr} \left( r^{D-2} \, r^{1-D/2}J_{\nu(D,l)}( \omega r) \right)
\right\vert_{r=a}&=&0, \mbox{ for region I TM-modes}, \label{condTMI}
\end{eqnarray}
Clearly, when a particular solution is of TE type, $\omega a$ has to be
a nonvanishng zero of a Bessel function $J_{\nu(D,l)}$, for some $l$.
\req{condTMI} can also be written as
\beq
( D/2-1 ) J_{\nu(D,l)}(k)+k J_{\nu(D,l)}'(k)=0,
\ k\equiv \omega a ,
\label{Rocond}
\eeq
which is a Robin (or standard homogeneous)
condition with
relative
coefficients $( D/2-1, 1)$.
If one keeps in mind the Maxwell equations, all this is valid
for $D=3$ only but, when regarded as just two massless
scalar fields
obeying the Klein-Gordon equation and subject to Dirichlet and
(a specific) Robin b.c., it holds for general $D$\footnote{
 The possible relevance of the space dimension as a
 perturbation parameter was noted in \cite{Be}.}.
The same boundary conditions apply to the external (region II)
solutions and we shall come back later to this question.

What is more, taking advantage of the $D=3$ duality
$\vec{E} \leftrightarrow \vec{B}$ between e.m. and colour gauge fields
one realizes that, up to a global factor equal to the number of
SU($N_c$) degrees of freedom,
the present set-up is equivalent to a bag model
for linear QCD\footnote{by `linear' we mean that the nonlinear
$f^{abc}A^{\mu}_a A^{\nu}_b$ term is omitted from the field-strength
tensor, as usual in such approaches.}
including only internal gluon modes.

\subsection{Neutral and perfectly conducting circular wire in $D=2$}

Two-dimensional spaces can be of interest since they are the
scenario for systems such as quantum billiards\cite{Ste,BeRo,It} or
anyons \cite{any}.
In particular, the electromagnetic--Chern-Simons Casimir effect in a
$(2+1)$-dimensional spacetime has already been
studied for parallel conducting lines in \cite{MN1} and for
a circle in \cite{MN2}.
What is explained in this subsection might also be obtained by the
formalism set up in sects I and II of \cite{MN2} when the Chern-Simmons
part is absent.
The components of the e.m. tensor in a $(2+1)$-dimensional spacetime
are specified in terms of its electric and magnetic fields
\[
F_{\mu \nu} =
\left( \begin{array}{ccc}
0 & E_1 & E_2 \\
-E_1 & 0 & B  \\
-E_2 & -B & 0
\end{array}
\right) .
\]
In $D=2$ there is room for only one $\vec{B}$-component. Actually
the magnetic field has become a scalar rather than a vector.
We impose the Maxwell equations in the vacuum
(absence of charges and current density)
\beq
 \partial_{\mu} F^{\mu,\nu} = 0
\label{first}
\eeq
\beq
 \partial_{\alpha} F^{*\ \alpha} = 0 ,
\label{second}
\eeq
where
$F^{*\ \alpha} = \frac{1}{2} \epsilon ^{\mu \nu \alpha} F_{\mu \nu}$
is the dual tensor
to the e.m. field. More explicitly we have $F^{*\ 0}=B$ $F^{*\ 1}=-E_2$
and $F^{*\ 2}=E_1$.
 From (\ref{second}) we get
\beq
\rot E = \dot{B} , \label{rot}
\eeq
where we understand that $\rot E \equiv \partial_1 E_2 - \partial_2 E_1$.
 From equation (\ref{first}) we draw
\bea
\div{\vec E}&=& 0  \label{div} \\
\dot{E}_1 + \partial_2 B & = & 0 \label{nu=1} \\
\dot{E}_2 - \partial_1 B & = & 0 \label{nu=2} .
\eea
\req{rot}-\req{nu=2} are the complete set of fundamental equations.
Taking the time-derivative of \req{rot}, $\partial_2$ of \req{nu=1},
$\partial_1$ of \req{nu=2} and combining the results, one finds
$\nabla^2 B - \ddot{B}=0$ .i.e.
\beq \Box B=0 \label{KGB} \eeq
which shows that $B$ obeys a
Klein-Gordon equation.
Now,
we give the form of these expressions when the fields have a time
dependence of the form
$\vec{E}(\vec{x},t)= \vec{\epsilon}(\vec{x}) e^{-i \omega t}$
and $\vec{B}(\vec{x},t)= \vec{b}(\vec{x}) e^{-i \omega t}$
\bea
\rot\vec{\epsilon}&=& -i \omega b , \label{rot1} \\
\div\vec{\epsilon}&=& 0,  \label{div1} \\
i \omega \epsilon_1 - \partial_2 b & = & 0, \label{nu=11} \\
i \omega \epsilon_2 + \partial_1 b & = & 0. \label{nu=21}
\eea
 From this set of equations ---or straighforwardly from \req{KGB} and
the time dependence of $\vec{B}(\vec{x},t)$--- we get
that
$(\Delta + \omega ^2)b = 0$.
Once this equation is solved, the electric field is given by
equations \req{nu=11}, \req{nu=21}. The ensuing solution may be easily
seen to automatically satisfy expression \req{rot1}.
Now we arrive at the question of boundary conditions.
Ancient lore
tells us that the electric field must be orthogonal to the
surface of a perfect conductor; if the normal vector is given by
$(n_1,n_2)$, this condition takes the form $0=n_1 E_2 - n_2 E_1$, which,
using \req{nu=11}, \req{nu=21}
is equivalent to $0=\frac{\partial}{\partial \vec{n}} B$.
To sum up: the problem is reduced to that of a scalar field $B$
which satisfies the typical Helmholtz equation and satisfies
    Neumann boundary conditions.
In our notations, for $D=2$ one has
$\alpha(2)=0$ echoing the conversion of
the Robin b.c. into a purely     Neumann condition.

The reader may check that all the modes generate a non-zero total
charge on the boundary (which is given by the scalar product of the
electric field and the normal vector to the boundary).

\section{Internal modes}

Within these types of set-up,
zero-point energies emerge from vacuum mode-sums of the type
$\ds\hbar c \, {1 \over 2}\sum_n \omega_n$, and give rise
to the celebrated Casimir effect \cite{Cas,PMG}.
Note that the summation extends over all the $\omega_n$'s in the set of
eigenmodes. As a result, sums of this
sort do usually diverge and call for some regularization to make sense
of them.

At this point, we introduce the usual spectral zeta functions, which we
denote by
\beq
\zeta_{{\cal M}}(s)=\sum_n \omega_n^{-s}, \hspace{1cm}
\zeta_{{\cal M}\over \mu}(s)=
\sum_n \left( \omega_n\over \mu \right)^{-s},
\eeq
$\mu$ is an arbitrary scale with mass dimensions, often used to work
with dimensionless objects.
As they stand, these identities hold only for $\rep s > s_0$,
being $s_0$ a positive value given by the rightmost pole of
$\zeta_{{\cal M}}(s)$. However, such a function admits analytic
continuation to other values of $s$, in particular, to negative reals.

The finite part of the vacuum energy ---$E_C$---
can be found by combining
zeta-regularization of the mode-sum and
a principal part prescription from ref.\cite{BVW}.
Following that work, one may put
\begin{equation}
E_{C}(\mu)= {\rm PP}_{s\to -1}\left[
{1 \over 2} \hbar c \, \mu \, \zeta_{{\cal M}\over \mu}(s)
\right] ,
\label{PPP}
\end{equation}
where PP denotes principal part.
One should be aware that the whole, observable, physical energy
includes other terms which have to do with the couplings of the
bag Lagrangian --- see the discussion on this point in the same reference.
(From here on, we adopt the typical QFT units which make $\hbar=c=1$).
Obviously, for this procedure to work we must know how to obtain
the analytic continuation of $\zeta_{{\cal M}}(s)$
to ---at least--- a small part of the negative real axis.

In order to proceed, we shall introduce `partial-wave' zeta
functions for fixed values of the Bessel index $\nu$.
We define the zeta function for the zeros of
$J_{\nu}$ as (see also \cite{ELR,LR}).
\beq
\zeta_{\nu}^{{\rm I}, {\cal D}}(s)=\sum_{n=1}^{\infty} j_{\nu, n}^{-s} \, ,  \
\mbox{for $\rep s > 1$},
\label{zetanuSe}
\eeq
where $j_{\nu n}$ denotes the $n$th nonvanishing zero of $J_{\nu}$.
The ${{\rm I}, {\cal D}}$ label has been added as a reminder that this comes
from eigenmodes in region I dictated by Dirichlet-type b.c.\footnote{
In the mathematical literature, this object taken at even integer $s$
is sometimes referred to as the Rayleigh function \cite{Math}.}
(Discrete versions of the Bessel problem, their solutions and
associated zeta functions have also received some attention in
\cite{Kvit}).
Analogously, let
\beq
\zeta_{\nu}^{{\rm I}, {\cal R}}(s)=\sum_{n=1}^{\infty} k_{\nu, n}^{-s} \, ,  \
\mbox{for $\rep s > 1$},
\label{zetanuRoSe}
\eeq
with $k_{\nu n}$ denoting the $n$th solution of
eq. \req{Rocond} for a given $\nu$.

Reconsidering the same problem in $D$-dimensional space,
taking into account
the degeneracy $d(D,l)$ of each angular mode in $D$ dimensions, we
define the `complete' spherical zeta functions
\begin{equation}
\begin{array}{lllll}
\ds\zeta_{\cal M}^{{\rm I}, {\cal D}}(s)&=&
\ds a^s \sum_{l=l_{\rm min}}^{\infty} d(D,l)
\sum_{n=1}^{\infty} j_{ \nu(D,l), n}^{-s}
&=&\ds a^s \sum_{l=l_{\rm min}}^{\infty} d(D,l) \,
\zeta_{\nu(D,l)}^{{\rm I}, {\cal D}}(s) , \\
\ds\zeta_{\cal M}^{{\rm I}, {\cal R}}(s)&=&
\ds a^s \sum_{l=l_{\rm min}}^{\infty} d(D,l)
\sum_{n=1}^{\infty} k_{ \nu(D,l), n}^{-s}&=&
\ds a^s \sum_{l=l_{\rm min}}^{\infty} d(D,l) \,
\zeta_{\nu(D,l)}^{{\rm I}, {\cal R}}(s) .
\end{array}
\label{defzdDs}
\end{equation}
$l_{\rm min}$ is the minimum value of $l$ and, for gauge fields in $D=3$,
$l_{\rm min}=1$.
If we consider scalar fields, then $l_{\rm min}= 0$.

As for $d(D,l)$, we find its value in \cite{Vil} and put
\begin{equation}
\begin{array}{lll}
\ds d(D,l)&=&\ds (2l+D-2) { (l+D-3)! \over l! (D-2)! }
= {2 \over (D-2)!}\sum_{k=0}^{k_{\mbox{\scriptsize max}}(D)}
(-1)^k {\cal A}_k(D) \nu(D,l)^{D-2-2k},
\end{array}
\label{dDl}
\end{equation}
where
\begin{equation}
k_{\mbox{\scriptsize max}}(D)= \left\{
\begin{array}{ll}
\ds{D-3 \over 2}&\mbox{for odd $D \geq 3$}, \\
\ds{D\over 2} -2&\mbox{for odd $D \geq 4$},
\end{array}
\right.
\label{valkmax}
\end{equation}
and the form of the ${\cal A}_k(D)$'s can be read off
from \cite{CC} (see also \cite{RU}).

\subsection{Internal Dirichlet (TE) modes}

\subsubsection{`Partial-wave' zeta function}

By \req{PPP}, computing the Casimir energy calls for
the knowledge of the Bessel zeta functions \req{zetanuSe} at $s=-1$,
while
the complex domain where \req{zetanuSe} holds is bounded by $\rep s =1$.
Fortunately for us,
$\zeta_{\nu}^{{\rm I}, {\cal D}}(s)$ admits
an analytic continuation to other values of $s$.
What is more,
in refs. \cite{ELR} and \cite{LR} we showed how to obtain an
integral representation valid for $-1 < {\rm Re}\  s < 0$, which reads
\begin{equation}
\zeta_{\nu}^{{\rm I}, {\cal D}}(s)={s \over \pi} \sin{\pi s \over 2}
\int_0^{\infty} dx \, x^{-s-1}
\ln\left[ \sqrt{2\pi x} \, e^{-x} I_{\nu}(x) \right] ,  \  \
\mbox{for $-1 < {\rm Re}\  s < 0$}.
\label{zetanuID}
\end{equation}
Yet, we shall have to work out (\ref{zetanuID}) in order to obtain
an equivalent representation more amenable to numerical calculation.
The first step is to rescale $x \to \nu x$. Afterwards,
we will perform a subtraction procedure on the resulting expression.
The aim of that is the reduction of $\zeta_{\nu}^{{\rm I}, {\cal D}}(s)$
to some elementary functions of $s$ plus an integral, whose
integrand should be relatively easy to express in terms of
uniform asymptotic expansions\footnote{also called Debye expansions}
(u.a.e.'s).

The piece which we will subtract and add to the integrand of
\req{zetanuID} is
\begin{equation}
\begin{array}{cc}
\ds x^{-s-1}
\ln\left[ {\sqrt{x} \over (1+x^2)^{1/4}} e^{\nu( \eta(x)-x )} \right] =
x^{-s-1}&\ds\left[
\sigma^{{\rm I}, {\cal D}}_1
\ln{ (1+x^2)^{1/4} \over \sqrt{x} }
+ \sigma^{{\rm I}, {\cal D}}_2
\nu( \eta(x)-x )
\right], \\
&\ds
\sigma^{{\rm I}, {\cal D}}_1= -1, \hspace{1cm}
\sigma^{{\rm I}, {\cal D}}_2= +1.
\end{array}
\label{sbtermID}
\end{equation}
where
$\eta(x)$ is the function appearing
in the known u.a.e. of the Bessel function (see e.g.\cite{AS}):
\beq
\eta(x)= \sqrt{ 1+x^2 } + \ln{ x \over 1 +\sqrt{ 1+x^2 } } ,
\eeq
and $\sigma^{{\rm I}, {\cal D}}_1$, $\sigma^{{\rm I}, {\cal D}}_2$
have been introduced for future convenience.

The added term is then separately integrated
using the intermediate steps
\begin{equation}
\begin{array}{lll}
\ds\int_0^{\infty} dx \, x^{-s-1}
\ln{ (1+x^2)^{1/4} \over \sqrt{x} }&=&
\ds {\pi\over 4s \sin {\pi s \over 2}}, \\
\ds\int_0^{\infty} dx \, x^{-s-1} \, (\eta(x)-x)&=&\ds
\ds\int_0^{\infty} dx \, x^{-s-1}
\ln{ x \over 1 +\sqrt{ 1+x^2 } }
+ \int_0^{\infty} dx \ x^{-s-1} ( \sqrt{ 1+x^2 } -x) \\
&=&\ds
\ds{1 \over 2s} B\left( {s+1 \over 2}, -{s \over 2} \right)
+ 2^{s-1}\left[ B\left( {s+1 \over 2}, -s \right)
              +  B\left( {s+3 \over 2}, -s \right) \right] ,
\end{array}
\label{FirstIntegr}
\end{equation}
which lead to
\begin{equation}
\begin{array}{lll}
\ds\zeta_{\nu}^{{\rm I}, {\cal D}}(s)&=&\ds
{1 \over 4} \sigma^{{\rm I}, {\cal D}}_1 \nu^{-s} \\
&&\ds+\nu^{-s} {s \over \pi} \sin{\pi s \over 2}\left[
\sigma^{{\rm I}, {\cal D}}_2 \left\{
\ds{1 \over 2s} B\left( {s+1 \over 2}, -{s \over 2} \right)
+ 2^{s-1} B\left( {s+1 \over 2}, -s \right) \right.\right. \\
&&\hspace{9em}\ds\left. + 2^{s-1} B\left( {s+3 \over 2}, -s \right)
\right\} \nu \\
&&\hspace{6em}\ds \left. +\int_0^{\infty} dx \ x^{-s-1}
\ln\left[
{\cal L}^{{\rm I}, {\cal D}}(\nu, x)
\right] \right] ,
\end{array}
\label{leadto}
\end{equation}
where, as usual
$\ds B(x,y)\equiv {\Gamma(x) \Gamma(y) \over \Gamma(x+y)}$, and
\beq
{\cal L}^{{\rm I}, {\cal D}}(\nu, x) \equiv
\sqrt{2\pi\nu}(1+x^2)^{1/4} e^{-\nu\eta(x)} I_{\nu}(\nu x) .
\label{LnuxID}
\eeq
The advantage of this new representation is that the u.a.e.
of the $\ln$ argument has reduced to
\begin{equation}
{\cal L}^{{\rm I}, {\cal D}}(\nu, x)
\sim
1+\sum_{k=1}^{\infty}{ u_k(t(x)) \over \nu^k}, \ \
t(x)={1 \over \sqrt{1+x^2}},
\label{uae}
\end{equation}
where the $u_k$'s are known polynomials listed in books like
\cite{AS}.
Expression \req{leadto} is also handy by the way in which the
pole at $s=-1$ is exhibited. The singularity at this point
is caused by:

\ni i) the $B$ functions with one argument equal to ${s+1 \over 2}$

\ni ii) the large-$x$ behaviour of the integrand; by (\ref{uae}) one
has
\[
\ln\left[
{\cal L}^{{\rm I}, {\cal D}}(\nu, x)
\right]
={1 \over 8\nu}t(x)+O(t^2(x))
={1 \over 8\nu x}+O\left( 1\over x^2 \right)
\]
which asymptotically yields a logarithmic divergence.
In fact, after integrating it has the same look as i), since
\beq
\int_0^{\infty} dx \, x^{-s-1} \, t(x)^m
={1 \over 2}B\left( {s+m \over 2}, -{s \over 2} \right) ,
\label{inttm}
\eeq
in this case with $m=1$.

The calculation of the integral in (\ref{leadto})
can be mentally divided into two stages.
First:
{\it necessarily } one has to delete and separately add
the part responsible for the divergence in ii).
Second: to make things numerically easier,
and eventually extract some infinities which will appear
later in the `complete' zeta function,
it is convenient to keep
subtracting and adding more terms of the expansion of
$\ln[ {\cal L}^{{\rm I}, {\cal D}}(\nu, x) ]$

Thus, taking several terms in the series, we write the
$\ln$ function in (\ref{leadto}) as
\begin{equation}
\ln\left[
{\cal L}^{{\rm I}, {\cal D}}(\nu, x)
\right]
\sim
\ln\left[
1+\sum_{k \geq 1} { u_k(t(x)) \over \nu^k}
\right]
= \sum_{n \geq 1} { {\cal U}^{{\rm I}, {\cal D}}_n(t(x)) \over \nu^n },
\label{loguae}
\end{equation}
where
the ${\cal U}^{{\rm I}, {\cal D}}_n$'s are given by
\begin{equation}
\begin{array}{lll}
\ds{\cal U}^{{\rm I}, {\cal D}}_1(t)&=&\ds
{t\over 8} - {{5\,{t^3}}\over {24}}, \\
\ds{\cal U}^{{\rm I}, {\cal D}}_2(t)&=&\ds
{{{t^2}}\over {16}} - {{3\,{t^4}}\over 8} +
 {{5\,{t^6}}\over {16}}, \\
\ds{\cal U}^{{\rm I}, {\cal D}}_3(t)&=&\ds
{{25\,{t^3}}\over {384}} - {{531\,{t^5}}\over {640}} +
     {{221\,{t^7}}\over {128}} - {{1105\,{t^9}}\over {1152}}, \\
\ds{\cal U}^{{\rm I}, {\cal D}}_4(t)&=&\ds
     {{13\,{t^4}}\over {128}} -
     {{71\,{t^6}}\over {32}} +
     {{531\,{t^8}}\over {64}} -
     {{339\,{t^{10}}}\over {32}} +
     {{565\,{t^{12}}}\over {128}} , \\
&\vdots&
\end{array}
\label{U14}
\end{equation}
  From them, we form the quantities
\beq
\begin{array}{lll}
\ds {\cal J}^{{\rm I}, {\cal D}}_n(s)&\equiv &\ds \int_0^{\infty} dx \,
x^{-s-1} \, {\cal U}^{{\rm I}, {\cal D}}_n(t(x)).
\end{array}
\label{JsfromUs}
\eeq
The expressions for these ${\cal J}^{{\rm I}, {\cal D}}_n(s)$'s
are easily obtained from the ${\cal U}^{{\rm I}, {\cal D}}_n(t)$'s in
(\ref{U14}) by application of \req{inttm}, as a result
of which it is enough to make the replacement
\beq
\begin{array}{rcl}
\ds{\cal U}^{{\rm I}, {\cal D}}_n(t)&\to&
\ds{\cal J}^{{\rm I}, {\cal D}}_n(s) \\
t^m&\to&\ds{1 \over 2}B\left( {s+m \over 2}, -{s \over 2} \right)
\end{array}
\eeq
With this, we may write
\begin{equation}
\begin{array}{c}
\ds\int_0^{\infty} dx \, x^{-s-1}  \,
\ln\left[
{\cal L}^{{\rm I}, {\cal D}}(\nu, x)
\right]
= {\cal S}^{{\rm I}, {\cal D}}_N(s, \nu)
+ \sum_{n=1}^{N}{{\cal J}^{{\rm I}, {\cal D}}_n(s) \over \nu^n} , \\
\ds {\cal S}_N^{{\rm I}, {\cal D}}(s, \nu) \equiv
\ds\int_0^{\infty} dx \, x^{-s-1} \, \left\{
\ln\left[
{\cal L}^{{\rm I}, {\cal D}}(\nu, x)
\right]
-\sum_{n=1}^N { {\cal U}^{{\rm I}, {\cal D}}_n(t(x)) \over \nu^n }
\right\} , \\
\end{array}
\label{defSJ}
\end{equation}
the key point being that $S^{{\rm I}, {\cal D}}_N(s, \nu)$
is a {\it finite} integral at $s=-1$.
Of all the ${\cal J}^{{\rm I}, {\cal D}}_n(s)$'s,
${\cal J}^{{\rm I}, {\cal D}}_1(s)$ is special as it contains
the only contribution to the $s=-1$ pole coming from the integral
in \req{leadto}
(i.e. the outcome of the `first stage';
the rest is produced by the `second stage').
For this reason we change the notation to
\begin{equation}
{\cal J}^{{\rm I}, {\cal D}}_1(s)
={\rho^{{\rm I}, {\cal D}} \over 2}B\left( {s+1 \over 2}, -{s \over 2} \right)
+\bar{\cal J}^{{\rm I}, {\cal D}}_1(s) , \hspace{0.5cm}
\rho^{{\rm I}, {\cal D}}={1 \over 8}, \hspace{0.5cm}
\bar{\cal J}^{{\rm I}, {\cal D}}_1(s)
= -{5 \over 24} B\left( {s+3 \over 2}, -{s \over 2} \right) .
\label{rhoID}
\end{equation}
Here the symbol $\rho^{{\rm I}, {\cal D}}$ has been introduced
to make easier the expressions of the forthcoming cases.
Then, we have
\begin{equation}
\begin{array}{lll}
\ds\zeta_{\nu}^{{\rm I}, {\cal D}}(s)&=&\ds
{1 \over 4} \sigma^{{\rm I}, {\cal D}}_1 \nu^{-s} \\
&&\ds+\nu^{-s} {s \over \pi} \sin{\pi s \over 2}\left[
\sigma^{{\rm I}, {\cal D}}_2
\left\{
\ds{1 \over 2s} B\left( {s+1 \over 2}, -{s \over 2} \right)
+ 2^{s-1} B\left( {s+1 \over 2}, -s \right) \right.\right. \\
&&\ds\hspace{9em}\left. + 2^{s-1} B\left( {s+3 \over 2}, -s \right)
\right\} \nu \\
&&\hspace{6em}\ds +{\cal S}^{{\rm I}, {\cal D}}_N(s, \nu) \\
&&\hspace{6em}\ds\left. +{1 \over 2} \rho^{{\rm I}, {\cal D}}
B\left( {s+1 \over 2}, -{s \over 2} \right){1 \over \nu}
+\bar{\cal J}^{{\rm I}, {\cal D}}_1(s) {1 \over \nu}
+\sum_{n=2}^{N} {\cal J}^{{\rm I}, {\cal D}}_n(s) {1 \over \nu^n} \right] .
\end{array}
\label{myzetanus}
\end{equation}
Laurent-expanding near $s=-1$ we
find
$\ds
\zeta_{\nu}(s)={1-4\nu^2 \over 8\pi}{1 \over s+1}+O( (s+1)^0 ),
$
which gives the right value of the residue at this point
\cite{Ste,ELR,LR}.
Similarly,
by Taylor expanding close to $s=0$ we get
$\ds
\zeta_{\nu}(s)= -{1\over 2}\left( \nu +{1 \over 2} \right)+O(s),$
i.e., although we
started from a representation valid for $-1 < {\rm Re}\  s < 0$,
the correct value of $\zeta_{\nu}$ at $s=0$ is also recovered.

\subsubsection{`Complete' zeta function}

Next, we go on to the $D$-dimensional problem.
This is done by
inserting both (\ref{dDl}) and (\ref{myzetanus})
into the first formula of (\ref{defzdDs}).
Afterwards, we may trivially interchange the
$l$-summation and the $k$-summation, since the second
is finite.
Once we have done so, we are left with sums of the type
\begin{equation}
\sum_{l=l_{\rm min}}^{\infty}\nu(D,l)^{-z}=
\zeta_H\left( z, {D \over 2}-1+l_{\rm min} \right),
\end{equation}
(recall \req{defnul})
$\zeta_H$ denoting the Hurwitz zeta function. Thus, we arrive at
\begin{equation}
\begin{array}{l}
\ds\zeta_{\cal M}^{{\rm I}, {\cal D}}(s)=
\ds{2 a^s \over (D-2)!}\sum_{k=0}^{k_{\mbox{\scriptsize max}}(D)}
(-1)^k {\cal A}_k(D) \\
\ds
\times\left[ {1 \over 4}\sigma^{{\rm I}, {\cal D}}_1
\zeta_H\left( -D+2+2k+s, {D \over 2}-1+l_{\rm min} \right)
\right. \\
\hspace{2em}\ds +{s \over \pi} \sin{\pi s \over 2}\left\{
\sigma^{{\rm I}, {\cal D}}_2
\left(
{1 \over 2s} B\left( {s+1 \over 2}, -{s \over 2} \right)
+ 2^{s-1} B\left( {s+1 \over 2}, -s \right)
+ 2^{s-1} B\left( {s+3 \over 2}, -s \right)
\right) \right. \\
\hspace{8em}\ds\times
\zeta_H\left( -D+1+2k+s, {D \over 2}-1+l_{\rm min} \right) \\
\hspace{8em}\ds +\sum_{l=l_{\rm min}}^{\infty}
{\cal S}^{{\rm I}, {\cal D}}_N(s, \nu(D,l)) \nu(D,l)^{D-2-2k-s} \\
\hspace{8em}\ds +{1 \over 2} \rho^{{\rm I}, {\cal D}}
B\left( {s+1 \over 2}, -{s \over 2} \right)
\zeta_H\left( -D+3+2k+s, {D \over 2}-1+l_{\rm min} \right) \\
\hspace{8em}\ds
+\bar{\cal J}^{{\rm I}, {\cal D}}_1(s)
\zeta_H\left( -D+3+2k+s, {D \over 2}-1+l_{\rm min} \right) \\
\left.\left.\hspace{8em}\ds
+\sum_{n=2}^{N} {\cal J}^{{\rm I}, {\cal D}}_n(s)
\zeta_H\left( -D+2+2k+s+n, {D \over 2}-1+l_{\rm min} \right) \right\}
\right] .
\end{array}
\label{zetaBD}
\end{equation}
Examining again the origins of singularity at $s=-1$, we find

\ni a) The ones already present for $\zeta_{\nu}$, visible now as
$\sim B\left( {s+1 \over 2}, \dots \right)$.

\ni b) New pole contributions when the first argument of any of the
present Hurwitz zeta functions equals one,
including the terms with
$n$ and $k$ such that $n=D-2k$ in the last sum.

On the other hand, at $s=-1$, the series
$\ds\sum_{l=l_{\rm min}}^{\infty}
{\cal S}^{{\rm I}, {\cal D}}_N(s, \nu(D,l)) \, \nu(D,l)^{D-2-2k-s}$
appears to have a rather slow
numerical convergence but, since its net contribution is actually
little (and the larger $N$, the smaller it gets) in practice
it suffices to compute until an accuracy of few digits is achieved.

\subsubsection{Three-dimensional space}

As a result of \req{valkmax}, for $D=3$ one has
$k_{\mbox{\scriptsize max}}(3)=0$ and the
$k$-series in \req{zetaBD} reduces to the $k=0$-term,
with ${\cal A}_0=1$.
Further, we consider the description of the electromagnetic TE
modes and therefore set $l_{\rm min}=1$
(for an ordinary scalar field  $l_{\rm min}=0$).
We split
$\zeta_{\cal M}^{{\rm I}, {\cal D}}(s)$ into two pieces:
$\zeta_{{\cal M} 1}^{{\rm I}, {\cal D}}(s)$, containing the part
directly evaluable at $s=-1$,
and $\zeta_{{\cal M} 2}^{{\rm I}, {\cal D}}(s)$ which includes
the `mixed' terms with both singular and finite contributions. Then
\begin{equation}
\zeta_{\cal M}^{{\rm I}, {\cal D}}(s)
=\zeta_{{\cal M} 1}^{{\rm I}, {\cal D}}(s)
+\zeta_{{\cal M} 2}^{{\rm I}, {\cal D}}(s),
\label{BB1B2}
\end{equation}
where
\beq
\begin{array}{ll}
\ds\zeta_{{\cal M} 1}^{{\rm I}, {\cal D}}(s)=2a^s&\ds\left[
{1 \over 4}\sigma^{{\rm I}, {\cal D}}_1
\zeta_H\left( s-1, {3 \over 2} \right)  \right.  \\
&\ds +{s \over \pi} \sin{\pi s \over 2} \left\{
\sigma^{{\rm I}, {\cal D}}_2
2^{s-1} B\left( {s+3 \over 2} , -s \right)
\zeta_H\left( s-2, {3 \over 2} \right) \right. \\
&\ds\hspace{5em} +\sum_{l \geq 1}
{\cal S}^{{\rm I}, {\cal D}}_N\left( s, l+{1 \over 2} \right)
\left( l+{1 \over 2} \right)^{1-s} \\
&\ds\left.\left. \hspace{5em} +\bar{\cal J}^{{\rm I}, {\cal D}}_1(s)
\zeta_H\left( s, {3 \over 2} \right)
+\sum_{n=2 \atop n \neq 3}^N {\cal J}^{{\rm I}, {\cal D}}_n(s)
\zeta_H\left( n+s-1, {3 \over 2} \right) \right\} \right]
\end{array}
\label{zetaB1}
\eeq
and
\beq
\begin{array}{ll}
\ds\zeta_{{\cal M} 2}^{{\rm I}, {\cal D}}(s)=
2a^s \sin{\pi s \over 2}&\ds\left[
\sigma^{{\rm I}, {\cal D}}_2
\left\{ {1 \over 2s} B\left( {s+1 \over 2}, -{s \over 2} \right)
+2^{s-1} B\left( {s+1 \over 2}, -s \right) \right\}
\zeta_H\left( s-2, {3 \over 2} \right) \right. \\
&\ds+{1 \over 2} \rho^{{\rm I}, {\cal D}}
B\left( {s+1 \over 2}, -{s \over 2} \right)
\zeta_H\left( s, {3 \over 2} \right) \\
&\ds+\left. {\cal J}^{{\rm I}, {\cal D}}_3(s)
\zeta_H\left( s+2, {3 \over 2} \right) \right]
\end{array}
\label{zetaB2}
\eeq
Next, we will compute
$\zeta_{{\cal M} 1}^{{\rm I}, {\cal D}}(s)$ directly at $s=-1$
and Laurent-expand
$\zeta_{{\cal M} 2}^{{\rm I}, {\cal D}}(s)$ near $s=-1$.
With the specific values of
$\sigma^{{\rm I}, {\cal D}}_1$,  $\sigma^{{\rm I}, {\cal D}}_2$,
$\rho^{{\rm I}, {\cal D}}$ from \req{sbtermID}, \req{rhoID} and
taking $N=4$ subtractions, we get
\beq
\ds\zeta_{{\cal M} 1}^{{\rm I}, {\cal D}}(-1)={1 \over a \pi}\left[
{{719}\over {5760}} + {{1053\,\pi }\over {8192}} +
{{35\,{{\pi }^3}}\over {65536}}
+2 \sum_{l \geq 1}
{\cal S}^{{\rm I}, {\cal D}}_4\left( -1, l+{1 \over 2} \right)
\left( l+{1 \over 2} \right)^2 \right] ,
\eeq
where, after a numerical calculation (with $N=4$), we have found
\beq
\ds \sum_{l=1}^{\infty}{\cal S}^{{\rm I}, {\cal D}}_4
\left(-1, l+{1 \over 2} \right)
\left(l+{1 \over 2} \right)^2  \simeq 0.00024
\label{Snu2num}
\eeq
The part containing the singularities gives rise to the following
Laurent expansion
\beq
\begin{array}{ll}
\ds\zeta_{{\cal M} 2}^{{\rm I}, {\cal D}}(s)={1 \over a \pi}&\ds\left[
{2\over 315}  \left( {1 \over  s + 1 } +\ln a \right) \right. \\
&\ds +{18457 \over 60480}
     -{229 \over 20160}\gamma
     -{11 \over 672}\ln 2
     - \zeta_H'\left( -3, {3\over 2} \right)
      +{1 \over 4} \zeta_H'\left( -1, {3\over 2} \right) \\
&\ds\left. +O( s+1 ) \right] .
\end{array}
\eeq
Here the prime denotes derivative with respect to the first argument.
To complete the desired numerical evaluation, we still need the values
of $\ds\zeta_H '\left(-3, {3 \over 2} \right)$
and $\ds\zeta_H '\left(-1, {3 \over 2} \right)$,
which are found from the relation
$\ds\zeta_H\left( s, {3 \over 2} \right)=-2^s+(2^s-1)\zeta_R(s)$
----where $\zeta_R$ means the Riemann zeta function---
and from the knowledge of $\zeta_R'(-3)\simeq 0.005378$
and $\zeta_R'(-1)\simeq -0.165421$.

Equipped with all this, we are able to obtain
\begin{equation}
\zeta_{\cal M}^{{\rm I}, {\cal D}}(s)={1 \over a}
\left[
{2 \over 315\pi} \left( {1 \over s+1} +\ln a \right)
+0.27069
+O(s+1)
\right] .
\label{zetaMID}
\end{equation}
The residue of the pole at $s=-1$ is
$\ds
{\rm Res}\left[\zeta_{\cal M}^{{\rm I}, {\cal D}}(s), s=-1 \right]=
{ 2 \over 315\pi a} ,
$
Bearing in mind that a $1/2$-factor appears when going from the zeta
function for the Maxwell eigenmodes to that for the
Laplacian spectrum, we realize that this agrees with
the heat-kernel expansion of that
operator
in the Dirichlet case
(see e.g. \cite{Ke}).
The existence of this pole
indicates that the Casimir energy
under the present conditions is still
infinite after zeta funtion regularization, and its divergence
cannot be removed until the application of the PP prescription,
which in some sense amounts to renormalizing.
(
The issue of
infinities
for the bag model was considered, from the
cutoff viewpoint, in \cite{BeHa}.
A full discussion about the essence of divergences in Casimir
energy problems was supplied in \cite{Can}.
)

Repeating the calculation for $l_{\rm min}=0$ one obtains
\beq
\zeta_{\cal M}^{{\rm scal}, {\rm I}, {\cal D}}(s)={1 \over a}
\left[
{2 \over 315\pi} \left( {1 \over s+1} +\ln a \right)
+0.00889
+O(s+1)
\right]
\label{zetaScID}
\eeq
which corresponds to the internal modes of a {\it true} scalar field.
In fact, this value
may also be found by adding to \req{zetaMID} the
`partial wave' contribution of the $l=0$ mode alone, which is
\beq
{1 \over a}\zeta_{1/2}^{{\rm I}, {\cal D}}(-1)={\pi \over a}\zeta_R(-1)
= -{\pi \over 12 a}\simeq -0.26180{1 \over a}
\label{l0ID}
\eeq
(Note here that, since $J_{1/2}(x) \propto x^{-1/2}\sin x$,
$\zeta_{1/2}^{{\rm I}, {\cal D}}(s)=\pi^{-s}\zeta_R(s)$).
As one can check, the figures match. Moreover, the importance of
the lower-lying region of the spectrum is manifest, since the
$l=0$ part is almost as large as the rest, but with opposite sign.
An approximate calculation of \req{zetaScID} based on the heat-kernel
expansion of the Laplacian is given in app. B.

\subsubsection{Two-dimensional space}

One can now think of the circular wire problem in the plane.
Given that we are still dealing  with
a Dirichlet field, this part will not be needed
for the $D=2$ e.m. Casimir effect, and is done just for completeness.
Now $d(2,l)=2$ for $l>0$ and $d(2,0)=1$.
With the methods in the preceding subsubsections, i.e.
carefully taking \req{zetaBD} for $D=2$ and $l_{\rm min}= 1$,
we find:
\beq
\zeta_{\cal M}^{{\rm scal}, {\rm I}, {\cal D}}(s)={1 \over a}
\left[ -{16+ \pi \over 128\pi}
\left( {1 \over s+1} + \ln a \right)
+ 0.02436         + O(s+1) \right] .
\eeq
Observe that $\nu(2,l=0)=0$ stops us from making the
rescaling $x \to \nu x$ and applying u.a.e.'s.
A way of dealing with $l=0$ parts in $D=2$ will be shown later.

\subsection{Internal Robin (TM) modes}
Here we outline the changes when considering the eigenmodes
obeying the Robin b.c. \req{Rocond}.
The analogue of \req{zetanuID} for these conditions is
(see \cite{LR})
\begin{equation}
\zeta_{\nu}^{{\rm I}, {\cal R}}(s)={s \over \pi} \sin{\pi s \over 2}
\int_0^{\infty} dx \ x^{-s-1}
\ln\left[ \sqrt{2\pi \over x} e^{-x} \left(
x I_{\nu}'(x) + \alpha \, I_{\nu}(x) \right) \right] ,  \  \
\mbox{for $-1 < {\rm Re}\  s < 0$}
\label{reprzetanuIR}
\end{equation}
We shall take $\alpha(D)= D/2-1$
in accordance with \req{Rocond} and thus $\alpha(3)=1/2$.
The calculation will be performed by a subtraction procedure
similar to that for the TE modes but,
now, the piece which we remove and add to the integrand is
(instead of \req{sbtermID})
\begin{equation}
\begin{array}{cc}
\ds x^{-s-1}
\ln\left[ { (1+x^2)^{1/4} \over \sqrt{x} } e^{\nu( \eta(x)-x )} \right] =
x^{-s-1} &\ds\left[
\sigma^{{\rm I}, {\cal R}}_1
\ln{ (1+x^2)^{1/4} \over \sqrt{x} }
+ \sigma^{{\rm I}, {\cal R}}_2
\nu( \eta(x)-x )
\right], \\
&\ds\sigma^{{\rm I}, {\cal R}}_1= +1, \hspace{1cm}
   \sigma^{{\rm I}, {\cal R}}_2= +1.
\label{sbtermIR}
\end{array}
\end{equation}
The ensuing expression is similar to \req{leadto} but for these changes:

\ni a) The signs $\sigma^{{\rm I}, {\cal D}}_1, \sigma^{{\rm I}, {\cal D}}_2$
are replaced with $\sigma^{{\rm I}, {\cal R}}_1, \sigma^{{\rm I}, {\cal R}}_2$.

\ni b) Instead of \req{LnuxID}, for the present case one has
\[
\begin{array}{c}
\ds{\cal L}^{{\rm I}, {\cal R}}(\nu, x)=
\sqrt{2 \pi \nu}
{1 \over (1+x^2)^{1/4}} e^{-\nu\eta(x)} x I_{\nu}'(\nu x)
+{\alpha}{ 1 \over \nu \sqrt{1+x^2} }
\sqrt{2 \pi \nu} (1+x^2)^{1/4} e^{-\nu\eta(x)} I_{\nu}(\nu x)
\end{array}
\]
Then $\ln\left[ {\cal L}^{{\rm I}, {\cal R}}(\nu, x) \right]$
is expanded by taking advantage of
the u.a.e.'s of both $I_{\nu}'(\nu x)$ and $I_{\nu}(\nu x)$
(see \cite{AS} again)
\[
\ln\left[ {\cal L}^{{\rm I}, {\cal R}}(\nu, x)  \right] \sim
\ln\left[ 1+\sum_{k \geq 1}{v_k(t(x))\over \nu^k}
+{\alpha}{t(x) \over \nu}
\left( 1+\sum_{k \geq 1}{u_k(t(x))\over \nu^k} \right) \right]
= \sum_{n \geq 1} { {\cal U}_n^{{\rm I}, {\cal R}}(t(x)) \over \nu^n }
\]
where
\beq
\begin{array}{rcl}
\ds{\cal U}^{{\rm I}, {\cal R}}_1(t)&=&\ds
\left( -{3\over 8} + \alpha  \right) \,t + {{7\,{t^3}}\over {24}} , \\
\ds{\cal U}^{{\rm I}, {\cal R}}_2(t)&=&\ds
\left( -{3\over {16}} + {{\alpha }\over 2} -
      {{{{\alpha }^2}}\over 2} \right) \,{t^2} +
   \left( {5\over 8} - {{\alpha }\over 2} \right) \,{t^4} -
   {{7\,{t^6}}\over {16}} , \\
\ds{\cal U}^{{\rm I}, {\cal R}}_3(t)&=&\ds
\left( -{{21}\over {128}} + {{3\,\alpha }\over 8} -
      {{{{\alpha }^2}}\over 2} + {{{{\alpha }^3}}\over 3} \right) \,{t^3} +
   \left( {{869}\over {640}} - {{5\,\alpha }\over 4} +
      {{{{\alpha }^2}}\over 2} \right) \,{t^5} +
   \left( -{{315}\over {128}} + {{7\,\alpha }\over 8} \right) \,{t^7} \\
&&\ds +{{1463\,{t^9}}\over {1152}} , \\
\ds{\cal U}^{{\rm I}, {\cal R}}_4(t)&=&\ds
\left( -{{27}\over {128}} + {{3\,\alpha }\over 8} -
      {{{{\alpha }^2}}\over 2} + {{{{\alpha }^3}}\over 2} -
      {{{{\alpha }^4}}\over 4} \right) \,{t^4} +
   \left( {{109}\over {32}} - {{23\,\alpha }\over 8} +
      {{3\,{{\alpha }^2}}\over 2}
      - {{{{\alpha }^3}}\over 2} \right) \,{t^6} \\
&&\ds +\left( -{{733}\over {64}} + {{41\,\alpha }\over 8} -
              {{\alpha }^2} \right) \,{t^8} +
   \left( {{441}\over {32}} - {{21\,\alpha }\over 8}\right) \,{t^{10}} -
         {{707\,{t^{12}}}\over {128}} , \\
&\vdots&
\end{array}
\label{U14IR}
\eeq
These formulas parallel eqs. \req{loguae} and \req{U14}.
Looking at the linear
$t$-term in the first of \req{U14IR}
we realize that the analogue of
$\rho^{{\rm I}, {\cal D}}$ in the previous case (eq. \req{rhoID}) is
\beq
\rho^{{\rm I}, {\cal R}}= \alpha-{3 \over 8} .
\eeq
The expression for $\zeta_{\nu}^{{\rm I}, {\cal R}}(s)$ is
like \req{myzetanus}, replacing all the quantities with
${{\rm I}, {\cal D}}$-superscripts
by the corresponding ones obtained with
${{\rm I}, {\cal R}}$-superscripts.
When doing so, we must bear in mind that in the case under consideration
\beq
\begin{array}{c}
\ds{\cal S}^{{\rm I}, {\cal R}}_N(s, \nu)=
\int_0^{\infty} dx \, x^{-s-1} \, \left\{
\ln\left[ {\cal L}^{{\rm I}, {\cal R}}(\nu, x) \right]
-\sum_{n=1}^N { {\cal U}_n^{{\rm I}, {\cal R}}(t(x)) \over \nu^n }
\right\}
\label{SIRNsnu}
\end{array}
\eeq
(which is by construction a finite integral) and
\begin{equation}
\bar{\cal J}_1^{{\rm I}, {\cal R}}(s) =
-{5 \over 24} B\left( {s+3 \over 2}, -{s \over 2} \right) ,
\hspace{1cm}
{\cal J}^{{\rm I}, {\cal R}}_n(s) = \int_0^{\infty} dx \, x^{-s-1} \,
{\cal U}^{{\rm I}, {\cal R}}_n(t(x)), \ n \geq 2
\label{JnIRs}
\end{equation}
For studying the complete spherical problem, we must deal with
the $\zeta_{\cal M}^{{\rm I}, {\cal R}}(s)$ in the second line
of \req{defzdDs},
and apply the same formula  \req{dDl}
for the $d(D,l)$'s, obtaining a similar series
of Hurwitz functions, integrals, etc.
Later, we specialize it to $D=3$, (therefore $\alpha=1/2$) and
$l_{\rm min}=1$. Before going on, we
separate
$\ds
\zeta_{\cal M}^{{\rm I}, {\cal R}}(s)=
\zeta_{{\cal M} 1}^{{\rm I}, {\cal R}}(s)
+\zeta_{{\cal M} 2}^{{\rm I}, {\cal R}}(s)
$
following the same finiteness criterion as in the decomposition
\req{BB1B2}.
Then, the resulting
$\zeta_{{\cal M} 1}^{{\rm I}, {\cal R}}(s)$,
$\zeta_{{\cal M} 2}^{{\rm I}, {\cal R}}(s)$ are obtained from the
$\zeta_{{\cal M} 1}^{{\rm I}, {\cal D}}(s)$,
$\zeta_{{\cal M} 2}^{{\rm I}, {\cal D}}(s)$ in formulas
\req{zetaB1}, \req{zetaB2}
by just replacing all the objects having
${{\rm I}, {\cal D}}$-superscript by their counterparts with
${{\rm I}, {\cal R}}$-superscript.
We also need the numerical calculation
\beq
\ds \sum_{l=1}^{\infty}{\cal S}^{{\rm I}, {\cal R}}_4
\left(-1, l+{1 \over 2} \right)
\left(l+{1 \over 2} \right)^2  \simeq 0.00012
\eeq
(analogous to \req{Snu2num}, and done for $N=4$ too).
After adding $\zeta_{{\cal M} 1}^{{\rm I}, {\cal R}}(-1)$
and the
Laurent-expansion of $\zeta_{{\cal M} 2}^{{\rm I}, {\cal R}}(s)$ near
$s=-1$,
we find
\begin{equation}
\zeta_{\cal M}^{{\rm I}, {\cal R}}(s)={1 \over a}
\left[
{2 \over 45\pi} \left( {1 \over s+1} +\ln a \right)
-0.10285
+O(s+1)
\right] .
\label{zetaMIR}
\end{equation}
For a scalar field, one can either repeat the calculation with
$l_{\rm min}=0$ or separately add the contribution from this mode.
The second option is easy because,
using $J_{1/2}(x)\propto x^{-1/2}\sin x$, our Robin condition
for $\nu(3,0)=1/2$ just reads $x^{1/2} \cos x=0$.
Its nonvanishing solutions are
$\pi\left(n+ {1\over 2} \right), n=0,1,2, \dots$
and therefore
$\ds
\zeta_{\nu=1/2}^{{\rm I}, {\cal R}}(s)=
\pi^{-s}\zeta_H\left( s, {1 \over 2} \right) .
$
At $s=-1$ one has
\beq
{1 \over a}\pi\zeta_H\left(-1, {1 \over 2} \right)={\pi \over 24 a}
=0.13090{1 \over a}.
\label{l0IR}
\eeq
So, after adding this part,
\beq
\zeta_{\cal M}^{{\rm scal}, {\rm I}, {\cal R}}(s)={1 \over a}
\left[
{2 \over 315\pi} \left( {1 \over s+1} +\ln a \right)
+0.02805
+O(s+1)
\right] .
\label{zetaScIR}
\eeq

The situation $D=2$, $\alpha=0$ may be similarly studied,
arriving at the complete zeta
function for the     Neumann modes with $l_{\rm min}=1$:
\beq
\zeta_{\cal M}^{{\rm I}, {\cal N}}(s)=
\left.\zeta_{\cal M}^{{\rm I}, {\cal R}}(s)\right\vert_{\alpha=0}=
{1 \over a}\left[ {48-5\pi \over 128\pi}
\left( {1 \over s+1} +\ln a     \right)
+ 0.17883
+O(s+1)
\right] .
\eeq

\section{External modes}

As already commented,
the spectrum of modes ---$\omega$'s--- is determined by the effect of
the problem's conditions on the radial part of the wave solutions
(in QFT language we would say `the {\it field} solutions' ).
This applies both to the solutions in the interior (region I)
and to those outside the spherical surface (region II), which
shall be now considered.
There are at least two possible approaches, both of them leading to
the same result:

\ni{\bf 1.} First, we sketch the simplest one.
If we demand
that the external solutions behave
like outgoing radial waves $\sim e^{i\omega r}$ for $r \to \infty$,
(though the discussion might be repeated with asymptotically
ingoing partial waves as well)
their radial parts will be just
$\propto r^{1-D/2} H^{(1)}_{\nu(D,l)}(\omega r)$,
with $H^{(1)}_{\nu}$ denoting the first Hankel function
(ingoing waves would just be the complex conjugate,
thus involving $H^{(1) \ *}=H^{(2)}$).
Following this cue, we repeat the contour integration
procedure of ref. \cite{ELR}, which gave the representations
\req{zetanuID}, \req{reprzetanuIR}, but putting now
$H^{(1)}_{\nu}$ instead of $J_{\nu}$ at the outset.
Since the calculations have the same nature, we do not
go throught them here;
perhaps the only point worthy of remark is that
wherever in ref.\cite{LR} we took advantage of properties like
$$
\begin{array}{c}
\ds J_{\nu}(z) \sim \sqrt{ 2 \over \pi z}
\cos\left( z-\left( \nu+ {1\over 2} \right){\pi \over 2} \right) ,
\ |z| \gg 1, \ | \mbox{arg} \, z| < \pi, \\
\ds J_{\nu}( e^{i\pi/2} z )=
e^{i\nu\pi/2} I_{\nu}(z),
\ -\pi <  \mbox{arg} \,  z \leq {\pi \over 2} ,
\end{array}
$$
now we have to make use of
$$
\begin{array}{c}
\ds H^{(1)}_{\nu}(z) \sim \sqrt{ 2 \over \pi z}
e^{i \left( z-\left( \nu+ {1\over 2} \right){\pi \over 2} \right) } ,
\ |z| \gg 1, \ -\pi <  \mbox{arg} \, z < 2\pi, \\
\ds H^{(1)}_{\nu}( e^{i\pi/2} z )=
-i{ 2\over \pi} e^{-i\nu\pi/2} K_{\nu}(z),
\ -\pi < \mbox{arg} \, z \leq {\pi \over 2} ,
\end{array}
$$
(and their equivalents for conjugates and derivatives). Doing so,
one obtains formulas \req{zetanuIID} and \req{zetanuIIR} below.

\ni{\bf 2.} Another reasoning, physically more transparent,
is to imagine a larger sphere of radius $R$, enclosing the one
of radius $a$, on whose surface we impose conditions as well
(in the siprit of ref.\cite{Bo}).
Once their partial-wave zeta function has
been constructed, the $R\to\infty$ limit is obtained. Not only is
the result $R$-independent, but it coincides with the outcome of
the method 1 as well. The full process is explained in app. A.

Once the desired `partial-wave' zeta functions
$\zeta_{\nu}^{{\rm II}, {\cal D}}(s)$,
$\zeta_{\nu}^{{\rm II}, {\cal R}}(s)$
have been obtained by either method,
taking the Bessel index $\nu(D,l)$
and the degeneracy $d(D,l)$ of each $l$ in $D$ dimensions, we
construct the `complete' spherical zeta functions
\begin{equation}
\begin{array}{lll}
\ds\zeta_{\cal M}^{{\rm II}, {\cal D}}(s)&=&
\ds a^s \sum_{l=l_{\rm min}}^{\infty} d(D,l) \,
\zeta_{\nu(D,l)}^{{\rm II}, {\cal D}}(s) , \\
\ds\zeta_{\cal M}^{{\rm II}, {\cal R}}(s)&=&
\ds a^s \sum_{l=l_{\rm min}}^{\infty} d(D,l) \,
\zeta_{\nu(D,l)}^{{\rm II}, {\cal R}}(s) .
\end{array}
\label{defzdIIDR}
\end{equation}
When considering the e.m. field,
we will eventually set $l_{\rm min}= 1$ (for a scalar one, $l_{\rm min}= 0$).

\subsection{Dirichlet (TE) external modes}
The adequate representation of the
`partial-wave' zeta function analogous to \req{zetanuID} for
region II is
\begin{equation}
\zeta_{\nu}^{{\rm II}, {\cal D}}(s)={s \over \pi} \sin{\pi s \over 2}
\int_0^{\infty} dx \, x^{-s-1} \,
\ln\left[ \sqrt{2x \over \pi} \, e^{x} K_{\nu}(x) \right]  ,  \  \
\mbox{for $-1 < {\rm Re}\  s < 0$} .
\label{zetanuIID}
\end{equation}
After rescaling $x\to\nu x$, a subtraction procedure similar to that
for the internal modes will take place, but
now
the piece which we remove and add to the integrand is
\beq
\begin{array}{cc}
\ds x^{-s-1}\ln\left[
{\sqrt{x} \over (1+x^2)^{1/4} } e^{-\nu( \eta(x)-x )}
\right]
\end{array}
\label{sbtermIID}
\eeq
As compared to \req{sbtermID}, this amounts to a sign flip in $\nu$.
Therefore
\[
\sigma^{{\rm II}, {\cal D}}_1= -1, \hspace{1cm}
\sigma^{{\rm II}, {\cal D}}_2= -1.
\]
The ensuing expression is similar to \req{leadto} but with the above
$\sigma_1$, $\sigma_2$ instead of those, and the
function ${\cal L}^{{\rm I}, {\cal D}}(\nu, x)$
appearing there replaced now with
\[
{\cal L}^{{\rm II}, {\cal D}}(\nu, x)=
\sqrt{2 \nu\over \pi} \, (1+x^2)^{1/4} e^{\nu \eta(x)} K_{\nu}(\nu x) .
\]
$\ln\left[ {\cal L}^{{\rm II}, {\cal D}}(\nu, x) \right]$
is expanded by taking adavantage of
the u.a.e. of $K_{\nu}(\nu x)$ (see \cite{AS} again)
$$\ds
\ln\left[ {\cal L}^{{\rm II}, {\cal D}}(\nu, x)  \right] \sim
\ln\left[ 1+\sum_{k \geq 1}(-1)^k{u_k(t(x))\over \nu^k}
\right]
= \sum_{n \geq 1} { {\cal U}_n^{{\rm II}, {\cal D}}(t(x)) \over \nu^n } ,
$$
where, by obvious parity reasoning with respect to the
${{\rm I}, {\cal D}}$ case (compare the above expression with \req{loguae})
\beq
{\cal U}^{{\rm II}, {\cal D}}_n(t)=
(-1)^n {\cal U}^{{\rm I}, {\cal D}}_n(t) .
\label{parity}
\eeq
As a result of \req{parity}
\beq
\rho^{{\rm II}, {\cal D}}= - \rho^{{\rm I}, {\cal D}}=-{1 \over 8} .
\eeq
Hence, by virtue of
\req{JsfromUs} and of \req{parity},
\begin{equation}
\bar{\cal J}_1^{{\rm II}, {\cal D}}(s) =
- \bar{\cal J}_1^{{\rm I}, {\cal D}}(s),
\hspace{1cm}
{\cal J}^{{\rm II}, {\cal D}}_n(s) =
(-1)^n {\cal J}^{{\rm I}, {\cal D}}_n(s) ,
\ n \geq 2 .
\end{equation}
With these elements,
we can already construct the analogue of \req{myzetanus},
say $\zeta_{\nu}^{{\rm II}, {\cal D}}(s)$,
for the external Dirichlet modes by `superscript substitution'.
The  resulting expression contains
\beq
\ds{\cal S}^{{\rm II}, {\cal D}}_N(s, \nu)=
\int_0^{\infty} dx \, x^{-s-1} \, \left\{
\ln\left[
{\cal L}^{{\rm II}, {\cal D}}(\nu, x)
\right]
-\sum_{n=1}^N { {\cal U}_n^{{\rm II}, {\cal D}}(t(x)) \over \nu^n }
\right\}
\eeq
which is a finite integral.
When studying the complete spherical problem, we must handle
the $\zeta_{\cal M}^{{\rm II}, {\cal D}}(s)$ in the first line of
\req{defzdIIDR},
and apply again formula \req{dDl}
for the $d(D,l)$'s.
Then, the emerging complete zeta function
is like \req{zetaBD}, but
with all ${{\rm I}, {\cal D}}$-superscripts turned into
${{\rm II}, {\cal D}}$-superscripts.
After setting $D=3$, $l_{\rm min}=1$, we
Laurent-expand around $s=-1$,
finding that everything is of the same sort as in the internal
case, except for the new quantity
$\ds
\ds \sum_{l=1}^{\infty}{\cal S}^{{\rm II}, {\cal D}}_4
\left(-1, l+{1 \over 2} \right)
\left(l+{1 \over 2} \right)^2  \simeq -0.00054
$
(here calculated for $N=4$ too).
So, after gathering everything together one arrives at
\begin{equation}
\zeta_{\cal M}^{{\rm II}, {\cal D}}(s)={1 \over a}
\left[
-{2 \over 315\pi} \left( {1 \over s+1} +\ln a \right)
-0.00326
+O(s+1)
\right] .
\label{zetaMIID}
\end{equation}
Redoing the calculation for $l_{\rm min}=0$, corresponding
to a scalar field, we obtain
$\ds
\zeta_{\cal M}^{{\rm scal}, {\rm II}, {\cal D}}(s)=
\zeta_{\cal M}^{{\rm II}, {\cal D}}(s),
$
i.e. the $l=0$ mode is not contributing in this case.

When $D=2$, $l_{\rm min}=1$, we get
\beq
\zeta_{\cal M}^{{\rm II}, {\cal D}}(s)={1 \over a}
\left[ {16-\pi \over 128\pi}
\left( {1 \over s+1} + \ln a    \right)
+ 0.00501
+O(s+1)
\right] .
\eeq
Now, the net contribution from the $l=0$ Dirichlet mode in $D=2$
will be found by
joining the inner and outer partial wave zeta functions
\req{zetanuID}, \req{zetanuIID}.
This sum has the effect of cancelling the $s=-1$ divergences
as a result of which we can numerically integrate, obtaining:
\beq
{1 \over a}\left[
\zeta_{\nu=0}^{{\rm I}, {\cal D}}(-1)+
\zeta_{\nu=0}^{{\rm II}, {\cal D}}(-1)
\right]=
{1 \over a\pi}\int_0^{\infty} dx \, \ln[ -2x I_0 (x) K_0 (x) ]
=-0.02802 {1 \over a} .
\label{zetaDn0IpII}
\eeq
Actually, by a calculation based on a slightly different representation
of $\zeta_{\nu}^{{\rm I}, {\cal D}}(s)$ valid for $\nu=0$ \cite{LREcqb},
we know that the part of \req{zetaDn0IpII} coming from the internal
modes is
$$
{1 \over a}\zeta_{\nu=0}^{{\rm I}, {\cal D}}(s)=
{1 \over a}\left[ {1 \over 8\pi}{1 \over s+1} -0.01451
+O(s+1)
\right] .
$$

\subsection{Robin (TM) external modes}
Applying any of the above referred procedures we find the
`partial-wave' zeta function representation
\begin{equation}
\zeta_{\nu}^{{\rm II}, {\cal R}}(s)={s \over \pi} \sin{\pi s \over 2}
\int_0^{\infty} dx \, x^{-s-1} \,
\ln\left[ \sqrt{2 \over \pi x} \, e^{x}
(-x K_{\nu}'(x)-\alpha K_{\nu}(x) ) \right]  ,  \  \
\mbox{for $-1 < {\rm Re}\  s < 0$} .
\label{zetanuIIR}
\end{equation}
The subtracted part will be
\begin{equation}
\begin{array}{cc}
\ds x^{-s-1}
\ln\left[ { (1+x^2)^{1/4} \over \sqrt{x} } e^{-\nu( \eta(x)-x )} \right]
\end{array}
\end{equation}
Thus, paralleling \req{sbtermIR},
\[
\sigma^{{\rm II}, {\cal R}}_1= +1, \hspace{1cm}
\sigma^{{\rm II}, {\cal R}}_2= -1.
\]
With respect to the situation described as ${{\rm I}, {\cal R}}$,
this amounts to a sign change in $\nu$
and the corresponding parity
reasonings apply everywhere (
as when going from ${{\rm I}, {\cal D}}$ to ${{\rm II}, {\cal D}}$).
In the present case we have to deal with
\beq
\begin{array}{c}
\ds{\cal L}^{{\rm II}, {\cal R}}(\nu,x)=
-\sqrt{2 \nu \over \pi}
{1 \over (1+x^2)^{1/4}} e^{\nu\eta(x)} x K_{\nu}'(\nu x)
-{\alpha}{ 1 \over \nu \sqrt{1+x^2} }
\sqrt{2 \nu \over \pi} (1+x^2)^{1/4} e^{\nu\eta(x)} K_{\nu}(\nu x) , \\
\ds{\cal S}^{{\rm II}, {\cal R}}_N(s, \nu)=
\int_0^{\infty} dx \, x^{-s-1} \, \left\{
\ln\left[ {\cal L}^{{\rm II}, {\cal R}}(\nu,x) \right]
-\sum_{n=1}^N { {\cal U}_n^{{\rm II}, {\cal R}}(t(x)) \over \nu^n }
\right\} ,
\end{array}
\eeq
where $
{\cal U}_n^{{\rm II} , {\cal R}}(t) =
(-1)^n {\cal U}_n^{{\rm I}, {\cal R}}(t)
$
and therefore
$\rho^{{\rm II} , {\cal R}}= -\rho^{{\rm I}, {\cal R}}$,
$\bar{\cal J}_1^{{\rm II} , {\cal R}}(s)=
-\bar{\cal J}_1^{{\rm I} , {\cal R}}(s)$,
${\cal J}_n^{{\rm II} , {\cal R}}(s)=
(-1)^n{\cal J}_n^{{\rm I} , {\cal R}}(s)$, $n \ge 2$.

Then, we construct the complete zeta function, which we calculate
for $D=3, \alpha=1/2$.
Taking into account that (for $N=4$)
$
\ds \sum_{l=1}^{\infty}{\cal S}^{{\rm II}, {\cal R}}_4
\left(-1, l+{1 \over 2} \right)
\left(l+{1 \over 2} \right)^2  \simeq -0.00041 ,
$
we are able to write the Laurent expansion near $s=-1$:
\begin{equation}
\zeta_{\cal M}^{{\rm II}, {\cal R}}(s)={1 \over a}
\left[
-{2 \over 45\pi} \left( {1 \over s+1} +\ln a \right)
-0.07223
+O(s+1)
\right] .
\label{zetaMIIR}
\end{equation}
As for a possible $l=0$ contribution when considering
a scalar field we see, like in the external Dirichlet case, that
this mode changes nothing, i.e.
$\ds
\zeta_{\cal M}^{{\rm scal}, {\rm II}, {\cal R}}(s)=
\zeta_{\cal M}^{{\rm II}, {\cal R}}(s).
$

When $D=2$, $\alpha=0$,
\beq
\zeta_{\cal M}^{{\rm II}, {\cal N}}(s)=
\left.\zeta_{\cal M}^{{\rm II}, {\cal R}}(s)\right\vert_{\alpha=0}=
{1 \over a}\left[ -{48+5\pi \over 128\pi}
\left( {1 \over s+1} +\ln a     \right)
- 0.03804
+O(s+1)
\right] .
\eeq
As in the $D=2$ Dirichlet case,
the total contribution from the $l=0$ mode alone follows from
adding the inner and outer partial wave zeta functions
which are now \req{reprzetanuIR} and \req{zetanuIIR}.
Again, the $s=-1$ divergences cancel and we can
find, by numerical integration:
\beq
{1 \over a}\left[
\zeta_{\nu=0}^{{\rm I}, {\cal R}}(-1)+
\zeta_{\nu=0}^{{\rm II}, {\cal R}}(-1)
\right]_{\alpha=0}=
{1 \over a\pi}\int_0^{\infty} dx \, \ln[ -2x I_0'(x) K_0'(x) ]
=-0.50704 {1 \over a}
\label{D2Nl0}
\eeq

\section{Discussion}

\subsection{$D=3$}
One can consider the contribution to the e.m. Casimir energy coming
from the interior of the sphere
only, i.e. the zero-point energy of a photon bag
banning the existence of outer modes.
Taking into account  \req{PPP}
and the sum of \req{zetaMID} plus \req{zetaMIR} one finds
\begin{equation}
E_{\rm C}^{\rm e.m. I}(\mu)= {1 \over a}
\left[ {8 \over 315\pi} \ln(a\mu)+ 0.08392 \right] .
\label{resEcembag}
\end{equation}
The logarithmic term, depending on $\mu$, has to be viewed as
a remainder of the renormalization process implicit in the
prescription adopted \cite{BVW}.
The second article of refs. \cite{Mi} gives, for
the Casimir energy due to the vector gauge bosons in the
interior of such a bag
\beq
{1 \over a}
\left[ {8 \over 315 \pi}\ln{\delta \over 8}
+ 0.08984
\right] ,
\label{resMilton}
\eeq
found
by the `energy method'
of \cite{MiRaSc}. $\delta$ is a cutoff arising from the
non-coincidence in time of field points,
linkable to the nonzero `skin depth' of
a real ---and not purely mathematical--- surface.
Incidentally, Riemann zeta functions were also employed in part of
the calculations in \cite{Mi}, although the initial
regularization approach was essentially different from ours.
Therefore, altough they are within a 6.7 \%,
there is no deep reason why the finite parts in
\req{resEcembag} and \req{resMilton} should be equal, as
figures may vary by just changing the values of the different
cutoffs.
In a (fermionless) QCD context $\mu$ can arguably be related
to the momentum scale parameter $\Lambda_{\rm QCD}$
(see e.g. \cite{Mu} and \cite{BVW}).
Results for that model follow by just
bringing in an obvious
$(N_c^2-1)$-factor from the SU($N_c$) degrees of
freedom; thus, for $N_c=3$
the cutoff-independent part of the energy becomes
$\sim {0.7 \over a}$.
(When fermions are assumed to be massless, neglecting them is not too bad
an approximation,
since results in the second work of refs.\cite{Mi} showed their
contribution as being one order of magnitude smaller).

Back to the e.m. case,
the same confining set-up ---without external modes--- in cubic cavities
(see \cite{Lu,AW}) yields a Casimir energy that amounts to $0.0916/L$,
where $L$ is the edge length.
The resemblance among this number and the finite parts of
\req{resEcembag} , \req{resMilton} happens to be striking, although
in view of the different details in the schemes leading to their
derivation one ought to be cautious before taking this point any further.
Joining \req{zetaScID} and
$\zeta_{\cal M}^{{\rm scal}, {\rm II}, {\cal D}}(s)$, which is equal to
\req{zetaMIID},
we shall find the net Casimir energy for a scalar field filling the
whole space and satisfying Dirichlet b.c. on the spherical surface.
{\it Whithout} having to apply \req{PPP}, one gets a finite
and scale-independent result, which reads
\begin{equation}
E_{\rm C}^{{\rm scal.} \ {\cal D}}= {1 \over a} \, 0.00282
\label{resEcscal}
\end{equation}
This finiteness is due to the cancellation of
both poles at $s=-1$ when adding up
internal and external contributions, and
may be put down to the
self-erasing of curvature-dependent infinities with
same size but opposite sign on each side of the surface.
\req{resEcscal} coincides with the energy value which would
yield the force (3.24) of ref. \cite{BeMi} for
the same physical situation, derived from a Green function approach.

Finally we consider an e.m. field in the whole space with
the sphere acting as a neutral and perfectly conducting boundary,
which corresponds to the sum of the four results
\req{zetaMID}, \req{zetaMIR}, \req{zetaMIID}, \req{zetaMIIR}.
On addition,
we realize that the poles cancel within the ${\cal D}$-pair
and within the ${\cal R}$-pair separately, rendering
the PP prescription in \req{PPP} unnecessary as we are left with
a finite and scale-independent value, namely
\begin{equation}
E_{\rm C}^{\rm e.m.}
={1 \over 2a} \, 0.09235
\label{resEcem}
\end{equation}
that coincides with the celebrated figure of \cite{Bo,BaDu,MiRaSc}.

Let's take another glance at all the results for $E_{\rm C}$
{\it prior} to the application of PP.
The outcome is summarized in table \ref{tabD3}.
\begin{table}[htbp]
\begin{center}
\begin{tabular}{|c|c|c|}
\hline\hline
$D=3$&${\cal D}$irichlet&
${\cal R}$obin $(\alpha(3)=1/2)$ \\ \hline
$\stackrel{ l=0 }{ \mbox{\ssz (only region I contributes)}  }$&
${1 \over 2a}\left[ -{\pi \over 12}\right]$&
${1 \over 2a} \, {\pi \over 24}$ \\
\hline
$\{ l \geq 1 \}$ {\nsz region I}&
$\barc
{1 \over 2a}\left[ {2 \over 315} \left( {1 \over s+1} + \ln(a\mu) \right)
\right. \\
\left. + 0.27069 \right]
\ear $&
$\barc
{1 \over 2a}\left[ {2 \over 45} \left( {1 \over s+1} +\ln(a\mu) \right)
\right. \\
\left. - 0.10285 \right]
\ear $
\\ \hline
$\{ l \geq 1 \}$ {\nsz region II}&
$\barc
{1 \over 2a}\left[ -{2 \over 315} \left( {1 \over s+1} +\ln(a\mu)\right)
\right. \\
\left. - 0.00326 \right]
\ear $&
$\barc
{1 \over 2a}\left[ -{2 \over 45} \left( {1 \over s+1}+\ln(a\mu) \right)
\right. \\
\left. - 0.07223 \right]
\ear $
\\ \hline\hline
\end{tabular}
\end{center}
\scapt{Zero-point energy decomposition in terms of scalar
fields satisfying Dirichlet and Robin b.c.
on a spherical surface of radius $a$ in $D=3$.}
\label{tabD3}
\end{table}
This way one easily sees that the PP prescription is redundant
when any internal-external pair is added up, which may be envisaged
as the above commented
surface divergence cancellation.
An analysis of heat-kernel coefficients allows one to realize that
these singularities are odd ---and therefore of
opposite sign on each side--- when the space dimension
is an odd number.
In zeta regularization
this is no longer so for curved surfaces in even $D$
(see \cite{BVW,ER-IJMPA1} --- notice also that the infinities in
\cite{BeMi} confirm this observation).
This fact has to be faced, in particular, in $D=2$.

\subsection{$D=2$}
Although we have argued that the e.m. problem in $D=2$
reduces to a     Neumann field,
for completeness the analogous figures
associated to a Dirichlet field were also calculated, and all
the values obtained have been listed in table \ref{tabD2}.
\begin{table}[htbp]
\begin{center}
\begin{tabular}{|c|c|c|}
\hline\hline
$D=2$&${\cal D}$irichlet&${\cal N}$eumann $(\alpha(2)=0)$ \\ \hline
$\stackrel{l=0}{ \mbox{\ssz (regions I + II)} }$&
${1 \over 2a}\left[ -0.02802 \right]$&
${1 \over 2a}\left[ -0.50704 \right]$ \\ \hline
$\{ l\geq 1 \}$ region I&
$\barc
{1 \over 2a}\left[ -{16+ \pi \over 128\pi}
\left( {1 \over s+1} + \ln(a\mu) \right) \right. \\
\left. + 0.02436 \right]
\ear $&
$\barc
{1 \over 2a}\left[ {48-5\pi \over 128\pi}
\left( {1 \over s+1} +\ln(a\mu) \right) \right. \\
\left. + 0.17883 \right]
\ear $
\\ \hline
$\{l \geq 1 \}$ region II&
$\barc
{1 \over 2a}\left[ {16-\pi \over 128\pi}
\left( {1 \over s+1} + \ln(a\mu)\right) \right. \\
\left. + 0.00501 \right]
\ear $&
$\barc
{1 \over 2a}\left[ -{48+5\pi \over 128\pi}
\left( {1 \over s+1} + \ln(a\mu)\right) \right. \\
\left. - 0.03804 \right]
\ear $
\\ \hline\hline
\end{tabular}
\end{center}
\scapt{Zero-point energy decomposition in
scalar fields under Dirichlet and  Neumann conditions on
a circular line of radius $a$ in $D=2$}
\label{tabD2}
\end{table}
Adding up all the Neumann parts, which are in the second column, one
finds \begin{equation}
E_{\rm C}^{\rm e.m. }(\mu)= {1 \over a}
\left[ -{5 \over 128} \ln(a\mu)- 0.18312 \right] .
\label{resEccircle}
\end{equation}
This minus sign makes us think of the attractive force obtained for
a cylinder in \cite{RaMi} (although that refers to three-dimensional
space, the symmetry is the same).
Not even the sum of inner and outer parts cancels all
the singularities.
This is a consequence of
using zeta regularization with a curved boundary
in even space dimension \cite{BVW,ER-IJMPA1}.

Doing the same for a scalar Dirichlet field we have
\begin{equation}
E_{\rm C}^{{\rm scal.} \ {\cal D}}(\mu)= {1 \over a}
\left[ -{1 \over 128} \ln(a\mu)+ 0.00068 \right] .
\label{resEcscaDcircle}
\end{equation}
The coefficient of the logarithmic part agrees with
the first ref. of \cite{Sen} for the same set-up.
Since the regularization method in that paper was
frequency cutoff, this coefficient is all that one
should expect to coincide. About possible ambiguities coming
from the use of different regularization schemes, see e.g. the
comments in \cite{CoVaZe}.

A comparison with the results in ref. \cite{MN2} is now in order.
The contributions from the $l=0$ mode in the Dirichlet (scalar) and
Neumann (e.m.) cases agree with formulas (A13) and (3.5)
---respectively--- in that work.
However, after performing the sum for
infinite values of $l$, a divergent part
$-{5 \over 128 a}{1 \over s+1}$ ---formula \req{resEccircle} prior to
prescription---
comes into existence through the
pole of $\zeta_H(z,1)=\zeta_R(z)$ when $z$ equals one
(i.e. what we have called, below eq.\req{zetaBD}, a b-type
singularity). Now, this specific piece has survived the infinity
cancellations which take place when adding all the internal and
external parts.
Observing the parity of all our ${\cal J}_n(s)$'s
when going from internal to external parts,
it is not difficult to realize that this happens for even $D \geq 2$.

These poles were also detected in sect. III of ref. \cite{BeMi}
after using dimensional regularization
and employing Riemann zeta functions
in the last stage of the calculation.
Within our zeta-regularization context, this
divergence would invalidate the alleged reliability of the finite
estimates in ref. \cite{MN2} for the $l \neq 0$ mode contribution
(formulas (A16) and (3.9) of that paper, for scalar and e.m. cases,
respectively).
The existence of a singularity was there acknowledged, but it was
argued that it could be a `spurious' one.
In our own framework, that speculation seems to be more questionable.
By way of connecting results,
we show in app. C how to reobtain those estimates by
performing what might be called
deliberately `na{\"\i}ve' zeta regularization.

\subsection{Ending comments}

Canonical field quantization leads to operators
---e.g. the Hamiltonian---  with ill-defined
vacuum expectation values. Such troubles, due
to quantum fluctuations, are often
suppressed
by decree
removing them when no observable effects are expected.
The picture is different in the presence of external sources or
boundaries that, by breaking symmetries, render fluctuations
observable. This explains the longevity of Casimir's general concept
of vacuum energy, according to which the physical vacuum of quantum
fields must be determined including their constraints.

Full vacuum energies may contain nonobvious infinite pieces
originated in boundary surface tensions, but
our attention here has focused
on the parts responsible for Casimir forces, i.e. those containing
dependence on the relevant space parameter,
which may be finite in spite of a global energy singularity.
That is the way in which `finiteness' has to be qualified \cite{Can}.

Viewed as an evaluation method for the Casimir effect, zeta
function regularization had been relatively successful up to now
but it was leaving
some vague aftertaste of scepticism insofar as it was applied
to comparatively simple problems: parallel plates, hypercubes,
torii, hyperspheres as framework spaces (not as boundaries),
i.e. situations where eigenvalues are at worst polynomials in the
quantum numbers. Circular or spherical boundaries are beyond
this realm.
The unified approach of the present work has quickly
enabled us to recover

\begin{itemize}
\item three remarkable Casimir energy results involving a sphere:

\begin{enumerate}

\item e.m. field inside:
second row of table \ref{tabD3}, which gives
the coefficient for the logarithmic term
in formula \req{resEcembag} as in \cite{Mi}.
The closeness
of the remaining pieces may be in principle fortuitous
since they come from different regularizations
(but is numerically convenient for comparing scales).
That is why we should regard our finite part
as a      new  zeta-regularization result.

\item scalar Dirichlet field inside and outside:
first column in table \ref{tabD3}, producing
\req{resEcscal} like in ref. \cite{BeMi}.

\item e.m. field inside and outside:
second plus third rows in table \ref{tabD3}, that yield
\req{resEcem} coinciding with refs. \cite{Bo,BaDu,MiRaSc},

\end{enumerate}

\item and one concerning a circle:

\begin{enumerate}

\item scalar Dirichlet field inside and outside:
first column in table \ref{tabD2}, giving
\req{resEcscaDcircle} whose logarithm coefficent is the
same as in \cite{Sen}. The rest is a new result
of zeta function regularization.

\end{enumerate}
\end{itemize}

Furthermore, combining figures other findings emerge,
e.g. the sum of all the contributions for the $D=3$
Robin scalar field ---second column in table \ref{tabD3}--- gives
$\ds
E_{\rm C}^{{\rm scal.} \ {\cal R}}=-0.02209 {1 \over a} ,
$
of a larger order of magnitude than and opposite sign to its
Dirichlet counterpart \req{resEcem}.
Moreover, the  zeta-function regularized version of the e.m. Casimir
energy inside and outside a circle \req{resEccircle} is,
to our knowledge, another unreleased result.

We hope that the new answers provided
---together with the recovery of figures originally obtained after
considerable effort--- by this single
zeta function strike
can give the reader reasons to think that the versatility and
scope of this technique may be somewhat wider.

\appendix\section{Appendix: external modes}

\subsection{External Dirichlet modes}

Here we give a detailed proof of expression \req{zetanuIID}.
Our starting point is the defining
series of the zeta function for a scalar field existing between two
spheres. The inner one has radius $a$ and is assumed to be physical.
The outer one has radius $R$ and it is introduced for the sake of
convenience; eventually we shall take the $R \rightarrow \infty$ limit.
Taking into account that $\nu(D=3,l) = l+\frac{1}{2}$,
initially we take solutions with radial part
$A h_l^{(1)} (\lambda r)+B h_l^{(2)} (\lambda r)$,
where the $h_l$'s are spherical Hankel functions and
$A$, $B$
are coefficients whose relative value ---with respect to each other---
will be determined by
imposing our Dirichlet conditions on both surfaces,
i.e. at $r=a$ and $r=R$. Doing so, we are left with an
homogeneous linear system for $(A,B)$ and, by requiring its
compatibility, the equation
\[ f_{a,R} (\lambda) \equiv
h_l ^{(1)} (\lambda a) h_l ^{(2)} (\lambda R) -
h_l ^{(1)} (\lambda R) h_l ^{(2)} (\lambda a)  = 0 \]
follows.
Therefore we study the `partial-wave' zeta function
$\ds\zeta_{\nu}(z)=\sum_p \frac{1}{\lambda_{\nu,p}^z}$, where
$\lambda_{\nu,p}$ is the p-th zero of the function $f_{a,R} (\lambda)$.

We tread here along the same lane that was open in \cite{LR,ELR},
which
runs through
\beq
\zeta_{\nu} (z) = \frac{z}{2 \pi i} \int d \lambda \,\,\, \lambda ^{-z-1}
\ln \left( f_{a,R} (\lambda) \right) .
\label{start}
\eeq
In expression (\ref{start}) the integration contour winds counterclockwise
round the
zeros of $f_{a,R}$, which are real. As it stands, this gives
a representation for $\zeta_{\nu}$ valid if ${\rm Re}\ z > 1$. We slightly
modify our representation to
\beq
\zeta_{\nu} (z) = \frac{z}{2 \pi i} \int d \lambda \,\,\, \lambda ^{-z-1}
\ln \left( \frac{i}{\lambda} \frac{\nu a^n R^n}{R^{2 \nu}-a^{2 \nu}}
f_{a,R} (\lambda) \right) ,
\label{start1}
\eeq
which ensures that the argument of the logarithm goes to 1 when $\lambda$
approaches 0; this will be seen to be useful in the sequel.

Now we deform the contour in such a way that it is made up of three parts:
a straight line from $+i \infty$ to $i \epsilon$, an arch of radius $\epsilon$
connecting $i \epsilon$
to $-i \epsilon$, and a straight line from $-i \epsilon$ to $-i \infty$.
The courteous reader may care to see in \cite{AS} that
\beq
f_{a,R} (\lambda ) = e^{i\lambda a-i\lambda R} S_n \left( -i \lambda a \right)
S_n \left( i\lambda R \right) - e^{-i \lambda a + i\lambda R}
S_n \left( -i\lambda R \right) S_n \left( i\lambda a \right) ,
\eeq
where
$\ds S_n \left( z \right) = \sum_{k=0}^n \frac{(n+k) !}{k ! (n-k) !}
(2 z)^{-k}$.

With this in mind, let us try to find an analytic continuation to the
contribution from the upper straight line. We immediately see that the
dominant contribution in the argument of the logarithm comes from the
$e^{i\lambda a - i\lambda R}$ term. So, we separate this factor in
the logarithm and apply the property that the logarithm of a product
equals the sum of the logarithms of its factors. This leads to
\beq
\begin{array}{c}
\ds \zeta_{\nu,+}(z)=-\frac{z}{2 \pi i^{z+1}}\left\{
\left( R-a \right) \frac{\epsilon ^{1-z}}{z-1} \right. \\
\ds\left. +\int_{\epsilon}^{\infty} \frac{d \rho}{\rho ^{z+1}
} \ln \left[ \frac{1}{\rho} \frac{\nu a^n R^n}{R^{2\nu}-a^{2\nu}} \left(
S_n (\rho a) S_n (-\rho R) -
e^{-2\rho (R-a)} S_n (\rho R) S_n (-\rho a) \right)
\right] \right\} .
\end{array}
\eeq
This expression explicitly defines an analytic function for ${\rm Re}\ z > 0$.
There is an equivalent representation for $\zeta_{\nu,-}$ (the contribution
from the lower straight line). It reads
\beq
\begin{array}{c}
\ds \zeta_{\nu ,-} (z)=-\frac{z i^{z+1}}{2 \pi}\left\{
\left( R-a \right)
\frac{\epsilon ^{1-z}}{z-1} \right. \\
\ds\left. +\int_{\epsilon}^{\infty} \frac{d \rho}{\rho ^{z+1} }
\ln \left[ \frac{1}{\rho } \frac{\nu a^n R^n}{R^{2\nu }-a^{2\nu }}
\left( S_n (\rho a) S_n (-\rho R)-
e^{-2\rho (R-a)} S_n (\rho R) S_n (-\rho a) \right)
\right] \right\} .
\end{array}
\eeq
The complete $\zeta_{\nu}$ is given by the addition of these contributions
plus the integration along the small arch of radius $\epsilon$. In any case
we have now an explicit continuation valid for ${\rm Re}\ z > 0$. If we
restrict to the domain $0 < {\rm Re}\ z < 1$, and take the limit
$\epsilon \rightarrow 0$ we end up with
\beq
\zeta_{\nu} (z) = \frac{z}{\pi} \sin \left( \frac{\pi}{2} z\right) \int_0^
{\infty} \frac{d \rho}{\rho ^{z+1}}
\ln \left[ \frac{1}{\rho} \frac{\nu a^n R^n}{R^{2\nu}-a^{2\nu}} \left(
S_n (\rho a) S_n (-\rho R) - e^{-2\rho (R-a)} S_n (\rho R) S_n (-\rho a) \right)
\right] .
\eeq
By now it may seem unpleasant that the large-$R$ limit is ill-defined.
When we go on with the continuation process towards domains in the complex
set with negative real part it will be apparent that in those regions one
may perform this limit, which will allow us to extract physical results.

In order to go on we have to add and subtract a function that apes
the asymptotic behaviour of the integrand for large $\lambda$
and which may be analytically
integrated, this is the general procedure
which was set forth in \cite{ELR,LR}.
Nevertheless, before we do this, we proceed to further simplify our expressions.
First of all, we separate the integral into three parts:
\bea
\zeta_{\nu} (z) & = &
\frac{z}{\pi} \int_L ^{\infty} \frac{d \rho}{\rho ^{z+1}}\ln \left[
S_n (\rho a) S_n (-\rho R) - e^{- 2 \rho (R-a)} S_n (\rho R) S_n (-\rho a)
\right] \nn
&&+ \frac{z}{\pi} \int_L ^{\infty} \frac{d\rho}{\rho ^{z+1}} \ln
\left[ \frac{1}{\rho} \frac{\nu a^n R^n}{R^{2 \nu}-a^{2 \nu}}
\right] \nn
&&+ \frac{z}{\pi} \int_0 ^L \frac{d\rho}{\rho ^{z+1}} \left( \ln \left[
\frac{1}{\rho} \frac{\nu a^n R^n}{R^{2 \nu}-a^{2 \nu}}
\right] + \ln \left[
S_n (\rho a) S_n (-\rho R) - e^{- 2 \rho (R-a)} S_n (\rho R) S_n (-\rho a)
\right] \right) \nn
 & \equiv & \zeta_{\nu,1}(z) + \zeta_{\nu,2}(z) + \zeta_{\nu,3}(z),
\eea
where $L$ is any positive number. The second integration is trivial and
gives a meromorphic function with a unique pole at $z=0$. The third
integration directly defines an analytic function in ${\rm Re}\ z < 1$.
It is the first integration which calls for the special treatment
that has been overviewed above.
So, we write
\beq
\begin{array}{c}
\ds\zeta_{\nu,1} (z) = \frac{z}{\pi} \sin \left( \frac{\pi}{2} z \right)
\\
\ds\times\int_L^{\infty } \frac{d\rho }{\rho ^{z+1}}
\left( \ln \left[ \frac{
S_n (\rho a) S_n (-\rho R) - e^{- 2 \rho (R-a)} S_n (\rho R) S_n (-\rho a)
}{A_n (\rho a,\rho R)} \right] + \ln \left( A_n (\rho a,\rho R) \right)
\right) ,
\end{array}
\eeq
where $A_n (\rho a,\rho R)$ is a function which shares the asymptotic
behaviour for large $\rho$ with
\[
S_n (\rho a) S_n (-\rho R) - e^{- 2 \rho (R-a)} S_n (\rho R) S_n (-\rho a)
\]
in such a way that
\beq
\frac{z}{\pi} \sin \left( \frac{\pi}{2} z \right)
\int_L^{\infty} \frac{d\rho}{\rho^{z+1}} \ln \left[ \frac{
S_n (\rho a) S_n (-\rho R) - e^{- 2 \rho (R-a)} S_n (\rho R) S_n (-\rho a)
}{A_n (\rho a,\rho R)} \right]
\eeq
defines an analytic function for $ -z_o < {\rm Re}\ z $ (where $z_o$
is a positive number which can be made as large as we wish by dint of
complicating $A_n$),  and
\beq
\frac{z}{\pi} \sin \left( \frac{\pi}{2} z \right)
\int_L^{\infty} \frac{d\rho}{\rho^{z+1}} \ln A_n (\rho a,\rho R)
\eeq
is easily computed, giving an explicit meromorphic function.
Once this is achieved we have our much coveted analytic continuation
of $\zeta_{\nu}$ for ${\rm Re}\ z > -z_o$. In this setting, we may restrict
the function $\zeta_{\nu}$ to $-z_o < {\rm Re}\ z < 0$; in this domain
we have
\beq
\zeta_{\nu} (z)  =  \zeta_{\nu,1} (z) +
\frac{z}{\pi} \sin \left( \frac{\pi}{2} z \right)
\int_0^L \frac{d\rho}{\rho^{z+1}} \ln \left(
S_n (\rho a) S_n (-\rho R) - e^{- 2 \rho (R-a)} S_n (\rho R) S_n (-\rho a)
\right)
\eeq
as $\zeta_{\nu,2}$ happens to cancel one contribution from
$\zeta_{\nu,3}$. To sum up, we write ($-z_o < z < 0$)
\bea
\zeta_{\nu} (z)  & = &
\frac{z}{\pi} \sin \left( \frac{\pi}{2} z \right)
\int_L^{\infty} \frac{d\rho}{\rho^{z+1}} \ln \left[ \frac{
S_n (\rho a) S_n (-\rho R)
- e^{- 2 \rho (R-a)} S_n (\rho R) S_n (-\rho a) }
{A_n (\rho a,\rho R)} \right] \nn
&&+ \frac{z}{\pi} \sin \left( \frac{\pi}{2} z \right)
\int_0^L \frac{d\rho}{\rho^{z+1}} \ln \left(
S_n (\rho a) S_n (-\rho R) - e^{- 2 \rho (R-a)} S_n (\rho R) S_n (-\rho a)
\right) \nn
&&+\mbox{Analytic\ continuation\ of} \left[
\frac{z}{\pi} \sin \left( \frac{\pi}{2} z \right)
\int_L^{\infty} \frac{d\rho}{\rho^{z+1}} \ln A_n (\rho a,\rho R) \right] .
\label{unda}
\eea
We have specified that we should calculate the analytic continuation
in the last term, as the integral
\beq
\frac{z}{\pi} \sin \left( \frac{\pi}{2} z \right)
\int_L^{\infty} \frac{d\rho}{\rho^{z+1}} \ln A_n (\rho a,\rho R)
\label{aux}
\eeq
only exists for ${\rm Re}\ z > -1$. Now, if $A_n (\rho a,\rho R)$ also
satisfies that expression (\ref{aux}) exists in the domain
$0 < {\rm Re}\ z < 1$ and its analytic continuation may be easily
computed in the particular case that $L=0$, then expression
(\ref{unda}) may be simplified to
\bea
\zeta_{\nu} (z) & = &
\frac{z}{\pi} \sin \left( \frac{\pi}{2} z \right)
\int_0^{\infty} \frac{d\rho}{\rho^{z+1}} \ln \left[ \frac{
S_n (\rho a) S_n (-\rho R) - e^{- 2 \rho (R-a)} S_n (\rho R) S_n (-\rho a)
}{A_n (\rho a,\rho R)} \right]  \nn
& & +\mbox{Analytic\ continuation\ of} \left[
\frac{z}{\pi} \sin \left( \frac{\pi}{2} z \right)
\int_0^{\infty} \frac{d\rho}{\rho^{z+1}} \ln A_n (\rho a,\rho R) \right] .
\label{fundasim}
\eea
The sagacious reader will immediately recognize that this expression
admits a simple large-$R$ limit, which may be written as
\bea
\zeta_{\nu} (z) & = &
\frac{z}{\pi} \sin \left( \frac{\pi}{2} z \right)
\int_0^{\infty} \frac{d\rho}{\rho^{z+1}} \ln \left[ \frac{
S_n (\rho a) }{A_n (\rho a)} \right]  \nn
&&+ \mbox{Analytic\ continuation\ of} \left[
\frac{z}{\pi} \sin \left( \frac{\pi}{2} z \right)
\int_0^{\infty} \frac{d\rho}{\rho^{z+1}} \ln A_n (\rho a) \right] ,
\label{fundasimlim}
\eea
where $A_n (\rho a)$ is supposed to have the asymptotic behaviour
of $S_n (\rho a)$ to some order. Note that this limit has been
made possible when we have passed to regions with negative
real part.
To finish this digression, we cast expression (\ref{fundasimlim})
into a more standard form:
\bea
\zeta_{\nu} (z) & = &
\frac{z}{\pi} \sin \left( \frac{\pi}{2} z \right)
\int_0^{\infty} \frac{d\rho}{\rho^{z+1}} \ln \left[ \frac{
\sqrt{\frac{2 \rho a}{\pi}} e^{\rho a} K_{n+\frac{1}{2}} (\rho a) }
{A_n (\rho a)} \right] \nn
&&+ \mbox{Analytic\ continuation\ of} \left[
\frac{z}{\pi} \sin \left( \frac{\pi}{2} z \right)
\int_0^{\infty} \frac{d\rho}{\rho^{z+1}} \ln A_n (\rho a) \right] ,
\label{fundasimlim1}
\eea
If we only need a representation valid in $-1 < {\rm Re}\ z < 0$
we need not include any $A_n$ at all and the result is
\beq
\zeta_{\nu} (z)  =
\frac{z}{\pi} \sin \left( \frac{\pi}{2} z \right)
\int_0^{\infty} \frac{d\rho}{\rho^{z+1}} \ln \left(
\sqrt{\frac{2 \rho a}{\pi}} e^{\rho a} K_{n+\frac{1}{2}} (\rho a) \right)
\eeq
which, when rescaled to $a=1$, and for
$n+1/2=\nu(D=3,n)$, gives \req{zetanuIID}.

\subsection{External Robin modes}

We explain now how to arrive at the expression for Robin
boundary condition in the external region.

In this derivation we will keep in mind what we did for the
modes satisfying Dirichlet boundary conditions. So we shall
also consider the existence of a large spherical shell of
radius $R$ which will be taken to $\infty$ once we have performed
the analytic continuation to domains with $\rep z < 0$.

Note that we will be assuming here that the conditions that the modes satisfy
on the outer sphere are also of the same Robin type as those
which are imposed on the physical one. This is equivalent
to saying that the space is confined by a large sphere which is a perfect
conductor. In any case this particular choice of boundary
conditions on the outer sphere has no physical relevance in the
sense that when one has performed the analytic continuation and $R
\rightarrow \infty$, the results do not depend on it, for instance,
we might as well decide to impose
Dirichlet boundary conditions for both TE and TM
modes on this unphysical sphere, and the final result (\ref{perfi})
would not be changed at all.

Setting Robin conditions at $r=a$ and $r=R$ on the radial
part of the spherical wave solution, one is led to consider
the $\lambda_{\nu(3,l),p}$'s which are zeros of the function
\beq
f_{a,R} (\lambda) \equiv
\sigma_l ^{(1)} (\lambda a) \sigma_l ^{(2)} (\lambda R)
- \sigma_l ^{(1)} (\lambda R) \sigma_l ^{(2)} (\lambda a) ,
\label{faRR}
\eeq
where $\sigma_l^{(i)}(z) = h_n^{(i)}(z)+z\frac{d}{dz}h_n^{(i)}(z)$
and $\nu(3,l) = l+\frac{1}{2}$. In more detail one has
\[
\sigma_n^{(1)}(z) = i^{-n-2} e^{iz}\left( -S_n (-iz)+ S_n^{'}(-iz)\right)
\]
\[
\sigma_n^{(2)}(z) = i^{n+2} e^{-iz}\left( -S_n (iz)+ S_n^{'}(iz)\right) .
\]

As we did in the case of Dirichlet boundary conditions we first
confine our efforts
to the study of $\ds\zeta_{\nu}(z)=\sum_p \frac{1}{\lambda_{\nu,p}^z}$.
which is again
given by \req{start}, but now with the $f_{a,R}(\lambda)$ in eq.
\req{faRR}.
The integration contour winds counterclockwise
round the poles of $f_{a,R}$, which are real (this property
is easily drawn from the fact that these poles are
eigenvalues of a self-adjoint operator). Our complex integral
is a valid representation for $\zeta_{\nu}$ if ${\rm Re}\ z > 1$.
We slightly
modify it by inserting a harmless factor in the argument of the logarithm,
which is again contrived so as to make the log an infinitesimal
quantity when $\lambda $ approaches the origin of the complex plane
\beq
\zeta_{\nu} (z) = \frac{z}{2 \pi i} \int d \lambda \,\,\, \lambda ^{-z-1}
\ln \left( \frac{i}{\lambda} \frac{\nu \lambda a^{n+1} R^{n+1}}
{n(n+1)i (R^{2 \nu}-a^{2 \nu})}
f_{a,R} (\lambda) \right) ,
\label{startR1}
\eeq
this is valid if $n>0$.

Now we deform the contour the same way as we did before:
a straight line from $+i \infty$ to $i \epsilon$, an arch of radius $\epsilon$
connecting $i \epsilon$
to $-i \epsilon$, and a straight line from $-i \epsilon$ to $-i \infty$.
Now the procedure closely follows what we did in the case of TE modes.
We find three contributions:
$\zeta_{\nu,+}$, $\zeta_{\nu,-}$
and the contribution from the arch.
Both  $\zeta_{\nu,+}$, $\zeta_{\nu,-}$ are given by
\beq
\zeta_{\nu,+}(z)  =  \frac{z i^{-z-1}}{2\pi } \frac{\epsilon ^{1-z}}{1-z}
(R-a)
-\frac{z i^{-z-1}}{2 \pi} \int_{\epsilon }^{\infty }
\frac{d\rho }{\rho ^{z+1}} \ln \left[ \frac{\nu \rho }{n(n+1)}\frac{a^{n+1}
R^{n+1}}{R^{2\nu } -a^{2\nu }} e^{-\rho (R-a)} f_{a,R} (i\rho ) \right]
\nonumber
\eeq
\beq
\zeta_{\nu,-}(z)  =  \frac{z i^{z+1}}{2\pi } \frac{\epsilon ^{1-z}}{1-z}
(R-a)
-\frac{z i^{z+1}}{2 \pi} \int_{\epsilon }^{\infty }
\frac{d\rho }{\rho ^{z+1}} \ln \left[ \frac{\nu \rho }{n(n+1)}\frac{a^{n+1}
R^{n+1}}{R^{2\nu } -a^{2\nu }} e^{-\rho (R-a)} f_{a,R} (i\rho ) \right] .
\nonumber
\eeq
It is immediate that we have realized an analytic continuation for
$\rep z > 0$.

If we
restrict the domain to $0 < {\rm Re}\ z < 1$, and take the limit
$\epsilon \rightarrow 0$ we end up with (note that in this limit
the contribution from the arch is vanishingly small)
\beq
\zeta_{\nu} (z) = \frac{z}{\pi} \sin \left( \frac{\pi}{2} z\right) \int_0^
{\infty} \frac{d \rho}{\rho ^{z+1}}
\ln \left[ \frac{-\nu \rho a^{n+1} R^{n+1}}
{n(n+1)(R^{2\nu}-a^{2\nu})}
f_{a,R} (-i\rho ) e^{-\rho (R-a)} \right] .
\eeq

For the sake of clarity we give here
\bea
-e^{-(R-a)} f_{a,R}(-i\rho ) & = &
\left( S_n ( \rho a )-S^{'}_n ( \rho a )  \right)
\left( S_n ( -\rho R )-S^{'}_n ( -\rho R ) \right)  \nn
&& -e^{-2(R-a)} \left[ \left( S_n( \rho R )-S^{'}_n ( \rho R ) \right)
\left( S_n ( -\rho a )-S^{'}_n ( -\rho a ) \right) \right] .
\eea

Now the reader should have no difficulty to apply the same procedure
that we have explained in the case of external TE modes. In particular,
we will have that for any $z_o$ there is a proper function $A_n(\rho a,
\rho R)$ such that the analytic continuation of $\zeta_{\nu}$
for $-z_o < \rep z < 0$ is
given by
\bea
\zeta_{\nu}(z) & = &\mbox{Analytic continuation of }\left[
\frac{z}{\pi }\sin \left( \frac{\pi}{2} z \right) \int_0^{\infty}
\frac{d\rho}{\rho ^{z+1}} \ln A_n(\rho a,\rho R) \right]  \nn
&&+\frac{z}{\pi }\sin \left( \frac{\pi}{2} z \right) \int_0^{\infty}
\frac{d\rho }{\rho ^{z+1}} \ln \left[
\frac{-e^{-(R-a)} f_{a,R}(-i\rho )}{A_n(\rho a,\rho R)} \right] .
\label{rotllo}
\eea
The choice of function $A_n$ is not uniquely determined. By simple
inspection one sees that it would be enough a function such that
for large $\rho $, $A_n(\rho a,\rho R)$ has the same
(algebraic) asymptotic
behaviour (to some convenient order) as the product
\[
\left( S_n(\rho a)-S^{'}_n(\rho a)\right)
\left( S_n(-\rho R)-S^{'}_n(-\rho R)\right) .
\]
A proper choice might be of the form $A_n(\rho a,\rho R) =
A_n^{(1)}(\rho a) A_n^{(2)}(\rho R)$, where $A_n^{(1)}(\rho a)$ is an
algebraic function that asymptotically behaves like
\[
S_n(\rho a)-S^{'}_n(\rho a)
\]
and $A_n^{(2)}(\rho R)$ does the same job for
\[
S_n(-\rho R)-S^{'}_n(-\rho R) .
\]
This would suffice for our purposes. As $\lim_{R\rightarrow \infty}
A_n^{(2)}(\rho R) = 1$, which is easily concluded from its
asymptotic behaviour, we see that in the domain which we are
considering it is possible the $R\rightarrow \infty$ limit. This
limit leads to
\bea
\zeta_{\nu}(z) & = &\mbox{\rm Analytic continuation of }\left[
\frac{z}{\pi }\sin \left( \frac{\pi}{2} z \right) \int_0^{\infty}
\frac{d\rho}{\rho ^{z+1}} \ln A_n^{(1)}(\rho a) \right] \nn
&&+ \frac{z}{\pi }\sin \left( \frac{\pi}{2} z \right) \int_0^{\infty}
\frac{d\rho }{\rho ^{z+1}} \ln \left[
\frac{S_n(\rho a)-S_n^{'}(\rho a)}{A_n^{(1)}(\rho a)} \right] .
\label{perfi}
\eea

If only a continuation valid in $-1 < \rep z < 0$ is needed, then we simply
have
\bea
\zeta_{\nu} (z)&=& \frac{z}{\pi }
\sin \left( \frac{\pi}{2} z \right) \int_0^{\infty}
\frac{d\rho }{\rho ^{z+1}} \ln \left[
S_n(\rho a)-S_n^{'}(\rho a)\right]
\eea
and, for $a=1$,
\bea
\zeta_{\nu}(z)&=&\frac{z}{\pi}\sin \left( \frac{\pi }{2} z\right)
\int_0^{\infty}
\frac{d \rho}{\rho^{z+1}} \ln \left[ \sqrt{\frac{2}{\pi z}} e^z
\left( -z K_{\nu}^{'} (z) -\frac{1}{2} K_{\nu} (z) \right) \right] ,
\eea
which gives \req{zetanuIIR} after noting that $\alpha(D=3)=1/2$.

\section{Appendix: heat-kernel series approximation to the complete
spectral zeta function}

For the Laplacian $\Box$ (or any such operator satisfying some suitable
requirements) the spectral zeta function $\zeta_{\Box}(z)=\Tr \Box^{-z}$
is related to the heat kernel $Y_{\Box}(t)=\Tr e^{-t\Box}$
through the well-known Mellin transform
\beq
\zeta_{\Box}(z)={1 \over \Gamma(z)}
\int_0^{\infty} dt \, t^{z-1} \, Y_{\Box}(t) .
\eeq
In turn, the heat kernel has small-$t$ asymptotic expansion
\beq
Y_{\Box}(t) \sim {1 \over (4\pi t)^{D/2}}\sum_{k \geq 0} b_{k/2} t^{k/2}.
\label{Ysmallt}
\eeq
We split the integration domain into the intervals $[0, \tau ]$ and
$[\tau, \infty)$, where $\tau$ is such that \req{Ysmallt} holds for
$t \leq \tau$. Thus, one can use this series in the first part and
integrate, finding
\beq
\zeta_{\Box}(z)=
{1 \over \Gamma(z)}\left[ {1 \over (4\pi t)^{D/2}}
\sum_{k \geq 0} { b_{k/2} \, \tau^{z+k/2-D/2} \over z+k/2-D/2 }
+ \int_{\tau}^{\infty} dt \, t^{z-1} \, Y_{\Box}(t) \right] .
\eeq
In our case, where the boundary is a sphere, the values of the
$b_{k/2}$'s can be read off from \cite{BoKi} (up to $k=20$ if
necessary!). Since $\Box$ gives squares of e.m. modes, symbolically
$\Box={\cal M}^2$ and $\zeta_{{\cal M}}(s)= \zeta_{\Box}(s/2)$, which
shall be studied around $s=-1$. Thus, separating the singular part
at this point we can write
\beq
\zeta_{{\cal M}}(s)={r \over s+1} +p(\tau)+c(\tau) +O(s+1),
\eeq
where
\beq
\begin{array}{rcl}
r&=&\ds {2 b_{(D+1)/2} \over
(4\pi)^{D/2} \Gamma\left( -{1 \over 2} \right) } ,\\
p(\tau)&=&\ds{1 \over (4\pi)^{D/2} \Gamma\left( -{1 \over 2} \right) }
\left[ b_{(D+1)/2}(\ln\tau -\psi(-1/2))
+2 \sum_{k \geq 0 \atop k \neq D+1}
{ b_{k/2} {\tau}^{(k-D-1)/2} \over k-D-1 } \right] , \\
c(\tau)&=&\ds {1 \over (4\pi)^{D/2} \Gamma\left( -{1 \over 2} \right) }
\int_{\tau}^{\infty} dt \, t^{-3/2} \, Y_{\Box}(t) .
\end{array}
\eeq
 From heat-kernel properties, one also knows that for large $t$
\beq
Y_{\Box}(t) \sim e^{-\lambda_0 t} , \label{Ylarget}
\eeq
where $\lambda_0$ stands
for the smallest eigenvalue of $\Box$.
We wish to use this behaviour in order
to approximate the value of $c(\tau)$.

As an example, we take $D=3$ and a scalar Dirichlet field inside
the sphere. From the $b_2$ listed in e.g. \cite{BoKi}, we
find
\beq r = {2 \over 315 \pi} \eeq
as we already knew (\req{zetaScID}).
$\lambda_0$ will be the square of the smallest nonvanishing zero of
$J_{1/2}(x)$ which is $x=\pi$; therefore $\lambda_0=\pi^2$.
Now, a delicate question arises about how to choose $\tau$ large enough
so that the replacement \req{Ylarget} be sensible while
maintaining the validity of the small-$t$ expansion \req{Ysmallt} for
$t \in [0,\tau]$. The answer
will necessarily be a compromise between both requirements. Looking at
the integral $c(\tau)$, which we hope to keep small,
one considers the region near $t=\tau$ and realizes
that it would be desirable to have $\lambda_0 \tau \gg 1$.
Thus one arrives at
$\ds{1 \over \lambda_0} \ll \tau < 1. $
Since $\lambda_0=\pi^2$,
we deem $\tau=1/2$ as a fairly adequate choice. Then we seek an
approximation to $c(1/2)$, which is shown by means of table \ref{tabhk}.
\begin{table}[htbp]
\begin{center}
\begin{tabular}{r|r}
largest $k$&approx. $p(1/2)$ \\ \hline
 5&0.009230 \\
10&0.009331 \\
15&0.009333 \\ \hline
\end{tabular}
\end{center}
\scapt{Approximations to $p(1/2)$, depending on the largest $k$
included in the sum.}
\label{tabhk}
\end{table}
On the other hand, the approximate value of the integral is,
by \req{Ylarget},
$c(1/2) \simeq -0.0004$. As a result, the finite part of
$\zeta_{\cal M}^{{\rm I}, {\cal D}}(s)$ is
\beq p(1/2) +c(1/2) \simeq 0.0093 -0.0004 = 0.0089  \eeq
in good agreement with the finite part of \req{zetaScID}.

\section{Appendix: `na{\"\i}ve' zeta-function regularization}
Let's calculate again the e.m. case in $D=2$, but making now the
addition of internal and external modes before the angular momentum
summation. After adding up \req{reprzetanuIR} to \req{zetanuIIR}
for $\alpha=0$ (i.e. purely Neumann b.c.) and rescaling $x \to\nu x$,
we find
\beq
\begin{array}{lll}
\zeta_{\nu}^{{\cal N}}(s)&\equiv&\ds
\left[
\zeta_{\nu}^{{\rm I}, {\cal R}}(s)
+\zeta_{\nu}^{{\rm II}, {\cal R}}(s)
\right]_{\alpha=0} \\
&=&\ds {s \over \pi}\sin{\pi s \over 2} \, \nu^{-s} \,
\int_0^{\infty} dx \, x^{-s-1} \,
\ln\left[ -2 \, \nu x I_{\nu}'(\nu x) K_{\nu}'(\nu x) \right] .
\end{array}
\eeq
Making use of the u.a.e.'s for $I_{\nu}'(\nu x)$ and $K_{\nu}'(\nu x)$,
taking advantage of previous steps and of parity reasoning,
\beq
\ln\left[ -2\nu x I_{\nu}'(\nu x) K_{\nu}'(\nu x) \right] \sim
\ln{ 1 \over x t(x) }
+ \sum_{n\geq 1}
{ \left. 2 \, {\cal U}^{{\rm I}, {\cal R}}_{2n}(t(x)) \right\vert_{\alpha=0}
\over \nu^{2n} }
\label{C2}
\eeq
where $t(x)$ is the one in \req{uae} and
the ${\cal U}^{{\rm I}, {\cal R}}_{2n}$'s are given in \req{U14IR}.
Subtracting up to $n=N$ terms from the integrand, one may write
\beq
\zeta_{\nu}^{{\cal N}}(s)=
{s \over \pi}\sin{\pi s \over 2} \, \nu^{-s} \, \left[
\int_0^{\infty} dx \, x^{-s-1} \,
\ln{ (1+x^2)^{1/2} \over x }
+\sum_{n=1}^N
{ \left. 2 \, {\cal J}^{{\rm I}, {\cal R}}_{2n}(s) \right\vert_{\alpha=0}
\over \nu^{2n} }
+{\cal S}^{{\cal N}}_N(s; \nu)
\right] ,
\eeq
By \req{FirstIntegr}, we know that the above integral amounts to
$\ds 2 \cdot { \pi \over 4 s \sin{\pi s \over 2} }$
The ${\cal J}^{{\rm I}, {\cal R}}_{2n}$'s are already specified by
the integrals \req{JnIRs}. ${\cal S}^{{\cal N}}_N(s; \nu)$ denotes
the corresponding residual integral, in the style of \req{SIRNsnu},
containing the difference between the exact integrand and its $N$-term
approximation.

Taking into account the $D=2$ degeneracies and noting that $\nu(2,l)=l$,
the complete zeta function is put as
\beq
\zeta^{{\cal N}}_{\cal M}(s)= a^s\left[
\zeta_{0}^{{\cal N}}(s) + 2 \sum_{l=1}^{\infty}\zeta_{l}^{{\cal N}}(s)
\right] .
\eeq
The $l=0$ contribution to the Casimir energy is therefore
\beq
{1 \over 2a}\zeta_{0}^{{\cal N}}(-1)
={1 \over 2a}
\left[
\zeta_{0}^{{\rm I}, {\cal R}}(s)
+\zeta_{0}^{{\rm II}, {\cal R}}(s)
\right]_{\alpha=0, \ s=-1} = -{1 \over a} 0.25352
\eeq
(formula \req{D2Nl0}). This agrees with (3.5) in ref. \cite{MN2}.
On the other hand, the set of $l \neq 0$ modes yields a contribution
which amounts to the $s \to -1$ limit of the following quantity:
\beq
\begin{array}{c}
\ds {1 \over a}\sum_{l=1}^{\infty}\zeta_{l}^{{\cal N}}(s)=
    {1 \over a}\left[
{1 \over 2}\zeta_R(s)
+{s \over \pi}\sin{\pi s \over 2}\left\{
\sum_{n=1}^N 2 \, \zeta_R(2n+s)
\left. {\cal J}^{{\rm I}, {\cal R}}_{2n}(s) \right\vert_{\alpha=0}
+\sum_{l=1}^{\infty} {\cal S}^{{\cal N}}_N(s; l) \, l^{-s}
\right\}
\right] .
\end{array}
\label{Cnlsums}
\eeq
At $s=-1$ the $l$-sum causes no problem, as it is finite and
relatively small. However, the $n$-sum contains a divergence in its
$n=1$-term, coming from the pole of $\zeta_R$ when its argument equals
one. Bearing this in mind, we find the residue of \req{Cnlsums}
at $s=-1$ to be
$
\ds\left.{2 \over \pi}{\cal J}^{{\rm I}, {\cal R}}_{2}(-1)
\right\vert_{\alpha=0}
=-{5 \over 128}
$,
as given by \req{resEccircle} (up this point, we have just been doing
the same calculation but re-grouping things in a convenient way).
Now, if we deliberately ignore this singularity and throw away as
`corrections' the $l-$ and $n-$ sums in \req{Cnlsums}, we simply get
\beq
{1 \over 2a}\zeta_R(-1)=-{1 \over 24 a},
\eeq
which is precisely the `LT' (`leading term') contribution of the
$l \neq 0$ modes to the Casimir effect given in eq. (3.9) of ref.
\cite{MN2},
i.e. it is the part which comes from keeping just the first term on the
r.h.s. of \req{C2}. Of course, viewed from inside our zeta-regularization
context, such an approximation in unjustifiable, since the whole value
has a divergence attached which we dare not call `spurious'.

Doing the same with the scalar Dirichlet modes, one easily sees that
the $l=0$ part is one half of \req{zetaDn0IpII}, in agreement with
result (A13) of ref. \cite{MN2}, and
the $l\neq 0$ piece analogous to the above mentioned approximation is
\beq
-{1 \over 2a}\zeta_R(-1)= {1 \over 24 a},
\eeq
which amounts to the `LT' result (A16) in ref. \cite{MN2} and whose
validity rests on the same flimsy assumptions.

\vskip5ex
\noindent{\Large \bf Acknowledgements}
\vskip3ex
Highly valuable comments by Kim A. Milton are appreciated.
Iver Brevik, Klaus Kirsten and Andrei A. Kvitsinsky
are thanked for discussions.
The authors are also grateful to the referee of this article for
the improvements suggested.
S.L. acknowledges an FI grant from {\it Generalitat de Catalunya}.
A.R. is thankful to {\it Generalitat de Catalunya ---Comissionat per a
Universitats i Recerca}--- for a RED fellowship, and to {\it CIRIT} for
further support.



\begin{thebibliography}{[00]}

\bibitem{Cas}
H.B.G. Casimir,
{\it Proc. Kon. Ned. Akad. Wetenschap} {\bf 51} (1948) 793.

\bibitem{To}
D.J. Toms,
{\it Phys. Rev.} {\bf D 21} (1980) 928;
{\it Phys. Rev.} {\bf D 21} (1980) 2805.

\bibitem{AW}
J. Ambj{\o}rn and S. Wolfram, {\it Ann. Phys.} {\bf 147} (1983) 1.

\bibitem{PMG}
G. Plunien, B. M{\"u}ller and W. Greiner,
{\it Phys. Rep.} {\bf 134} (1986) 87.

\bibitem{Fo}
L.H. Ford, {\it Phys. Rev. } {\bf D36} (1988) 528.

\bibitem{BVW}
S.K. Blau, M. Visser and A. Wipf,
{\it Nucl. Phys. } {\bf B310} (1988) 163.

\bibitem{BrSkSo}
I. Brevik and G.H. Nyland,
{\it Ann. Phys. } {\bf 230} (1994) 321; \\
I. Brevik, H. Skurdal and R. Sollie,
{\it J. Phys. } {\bf A 27} (1994) 6853.

\bibitem{RoWi}
E. Robaschik and E. Wieczorek,
{\it Ann. Phys. } {\bf 236} (1994) 43.

\bibitem{Po}
C. Erbelein,
{\it J. Phys.} {\bf A 25} (1992) 3015, 3039; \\
P.A. Ma{\"\i}a Neto and S. Reynaud,
{\it Phys. Rev.} {\bf A 47} (1993) 1639; \\
Y. Pomeau,
{\it Europhys. Lett.} {\bf 27} (1994) 377.

\bibitem{BeMi}
C.M. Bender and K.A. Milton,
{\it Phys. Rev.} {\bf D 50} (1994) 6547.

\bibitem{Ne}
A.V. Nesteruk,
{\it Europhys. Lett. } {\bf 32 } (1995) 455.

\bibitem{DeCa}
D. Deutsch and P. Candelas,
{\it Phys. Rev. } {\bf D 20} (1979) 3063.

\bibitem{MiRaSc}
K.A. Milton, L.L. DeRaad Jr., and J. Schwinger, {\it Ann. Phys. }
(N.Y.) {\bf 115} (1978) 388.

\bibitem{BaDu}
R. Balian and B. Duplantier, {\it Ann. Phys. } (N.Y.)
{\bf 112} (1978) 165.

\bibitem{SeGi}
H.P. McKean Jr. and I.M. Singer,
{\it J. Diff. Geom.} {\bf 1} (1967) 43; \\
R. Seeley,
{\it Amer. J. Math.} {\bf 91} (1969) 889; \\
P.B. Gilkey,
{\it Proc. Symp. Pure Math.} {\bf 27} (1975) 265;
{\it J. Diff. Geom.} {\bf 10} (1975) 601;
{\it Adv. in Math.} {\bf 102} (1993) 129; \\
T.P. Branson and P.B. Gilkey,
{\it Commun. Partial Diff. Eq.} {\bf 15} (1990) 245; \\
B.S. De Witt,
{\it Phys. Rep. } {\bf 19} (1975) 295.

\bibitem{Ke}
G. Kennedy, {\it J. Phys. } {\bf A 11} (1978) L173.

\bibitem{DuOlPe}
B. Durhuus, P. Olesen and J.L. Petersen,
{\it Nucl. Phys. } {\bf B 198} (1982) 157.

\bibitem{BoKi}
M. Bordag, E. Elizalde and K. Kirsten,
hep-th/9503023, {\it J. Math. Phys.} accepted.

\bibitem{Do}
J.S. Dowker,
hep-th/9506042; hep-th/9508082, {\it Phys. Lett. }{\bf B} accepted.

\bibitem{Mi}
K.A. Milton, {\it Phys. Rev.} {\bf D 22} (1980) 1441;
{\it Phys. Rev.} {\bf D 27} (1983) 439; \\
{\it Ann. Phys. } (N.Y.) {\bf 150} (1983) 432.


\bibitem{BMDC}
L.S. Brown and G.J. MacLay, {\it Phys. Rev.} {\bf 184} (1969) 1272; \\
J.S. Dowker and R. Critchley, {\it Phys. Rev.} {\bf D 13} (1976) 3224.

\bibitem{Haw}
S.W. Hawking, {\it Commun. Math. Phys. } {\bf 55} (1978) 133.

\bibitem{BeRo} M.V. Berry and M. Robnik,
{\it J. Phys.} {\bf A 19} (1986) 649; \\
M.V. Berry,
{\it J. Phys.} {\bf A 19} (1986) 2281;
{\it J. Phys.} {\bf A 20} (1987) 2389.

\bibitem{It} C. Itzykson, P. Moussa and J.M. Luck,
{\it J. Phys.} {\bf A 19}, L111 (1986); \\
R. M. Ziff,
{\it J. Phys. } {\bf A 19} (1986) 3923.

\bibitem{Dow}
J.S. Dowker, {\it J. Phys.} {\bf A 11 } (1978) 2255;
{\it Phys. Rev.} {\bf D 29} (1984) 2773.

\bibitem{CW}
P. Candelas and S. Weinberg,
{\it Nucl. Phys. } {\bf B237} (1984) 397; \\
A. Chodos and E. Myers,
{\it Ann. Phys. } {\bf 156} (1984) 412;
{\it Phys. Rev. } {\bf D 31} (1985) 3064; \\
R. Kantowski and K.A. Milton,
{\it Phys. Rev. } {\bf D 35} (1987) 549;
{\it Phys. Rev. } {\bf D 36} (1987) 3712.

\bibitem{GV}
G. Ghika and M. Vi\c{s}inescu,
{\it Nuovo Cimento } {\bf 46 A} (1978) 25.

\bibitem{KuUe}
T. Kunimasa and K. Uehara,
{\it Nucl. Phys. } {\bf B279} (1987) 608.

\bibitem{EORBZ}
E. Elizalde, S.D. Odintsov, A. Romeo, A.A. Bytsenko and S. Zerbini,
{\it Zeta Regularization Techniques with Applications}, World Scientific
(1994).

\bibitem{ELR} E. Elizalde, S. Leseduarte and A. Romeo,
{\it J. Phys.} {\bf A 26} (1993)  2409.

\bibitem{LR} S. Leseduarte and A. Romeo,
{\it J. Phys.} {\bf A 27} (1994) 2483.

\bibitem{LREcqb}
A. Romeo,
{\it Scalar Casimir effect in a circular Aharonov-Bohm quantum billiard},
unpublished;  \\
S. Leseduarte and A. Romeo,
{\it Influence of an external magnetic flux on the zero-point energy of
a field subject to spherical boundary conditions}, in preparation.

\bibitem{Math}
G.N. Watson,
{\sl A treatise on the Theory of Bessel Functions}, 2nd edition,
Cambridge University Press, Cambridge (1944); \\
N. Kishore, {\it Proc. Amer. Math. Soc.} {\bf 14} (1963) 527; \\
E.C. Obi, {\it J. Math. Anal. Appl.} {\bf 52} (1975) 648; \\
J. Hawkins, {\it On a zeta function associated with Bessel's equation},
PhD Thesis, University of Illinois (1983); \\
K.B. Stolarsky, {\it Mathematika} {\bf 32} (1985) 96.

\bibitem{Ste} F. Steiner,
{\it Fortschr. Phys. }{\bf 35} (1987)  87.

\bibitem{any}
Y-H. Chen, F. Wilczek, E. Witten and B.I. Halperin,
{\it Int. J. Mod. Phys. }{\bf B3} (1989) 1001; \\
S. Deser, {\it Phys. Rev. Lett. } {\bf 64} (1990) 611; \\
L. Alberto, {\it Anyons}, Springer Verlag (1992).

\bibitem{MN1}
K.A. Milton and Y.J. Ng,
{\it Phys. Rev. } {\bf D 42} (1990) 2875.

\bibitem{MN2}
K.A. Milton and Y.J. Ng,
{\it Phys. Rev. } {\bf D 46} (1992) 842.

\bibitem{RU}
A. Romeo,
{\it Phys. Rev. } {\bf D 52} (1995) 7308.

\bibitem{Bar}
A.O. Barvinsky, A. Yu. Kamenshchik and I.P. Karmazin,
{\it Ann. Phys.} {\bf 219} (1992) 201.

\bibitem{SlFr} J.C. Slater and N.H. Frank,
{\it Electromagnestism}, Dover, New York (1969).

\bibitem{Bo}
T.H. Boyer, {\it Phys. Rev. } {\bf 174} (1968) 174.

\bibitem{Be}
C.M. Bender, S. Boettcher and L. Lipatov, {\it Phys. Rev. Lett. }
{\bf 168} (1992) 3764;
{\it Phys. Rev. } {\bf D 146} (1992) 5557; \\
C.M. Bender and S. Boettcher, {\it J. Math. Phys.} {\bf 135} (1994) 1914;
{\it Phys. Rev. } {\bf D 148} (1993) 4919.


\bibitem{Kvit}
A.A. Kvitsinsky, {\it J. Phys. } {\bf A 28} (1995) 1753;
{\it J. Math. Anal. Appl.}, {\bf 196} (1995) 947.

\bibitem{Vil}
N. Ja. Vilenkin,
{\sl Fonctions sp{\'e}ciales et th{\'e}orie de la r{\'e}presentation des
groups}, Dunod, Paris (1969).

\bibitem{CC}
A. Cappelli and A. Coste, {\it Nucl. Phys.} {\bf B314} (1989) 707.

\bibitem{AS}
M. Abramowitz and I.A. Stegun, {\it Handbook of Mathematical Functions},
Dover, New York (1972).

\bibitem{BeHa}
C.M. Bender and P. Hays, {\it Phys. Rev.} {\bf D 14} (1976) 2622.

\bibitem{Can}
P. Candelas, {\it Ann. Phys. } (N.Y.) {\bf 143} (1982) 241.

\bibitem{Mu}
T. Muta, {\it Foundations of Quantum Chromodynamics}, World Scientific
(1987).

\bibitem{Lu}
W. Lukosz, {\it Physica} {\bf 56} (1971) 109.

\bibitem{Sen}
S. Sen, {\it J. Math. Phys.} {\bf 22} (1981) 2968; {\bf 25} (1984) 2000.

\bibitem{ER-IJMPA1}
E. Elizalde and A. Romeo, {\it Int. J. Mod. Phys. } {\bf A} (1990) 1653.

\bibitem{CoVaZe}
G. Cognola, L. Vanzo and S. Zerbini,
{\it J. Math. Phys. } {\bf 33} (1992) 222.

\bibitem{RaMi}
L.L. De Raad and K.A. Milton, {\it Ann. Phys. } {\bf 136} (1981) 229.


\end{thebibliography}
\end{document}